\documentclass[aps,prb,reprint,showpacs,superscriptaddress]{revtex4-1}

\usepackage[english]{babel}
\usepackage[utf8]{inputenc}
\usepackage{amsmath} 
\usepackage{bm}
\usepackage{amssymb}
\usepackage{graphicx}
\usepackage{color}
\usepackage[squaren,Gray]{SIunits}
\usepackage[version=4]{mhchem}
\usepackage[colorlinks=true, pdfstartview=FitV, linkcolor=blue, citecolor=blue, urlcolor=blue]{hyperref}
\usepackage{xspace}
\usepackage{soul} 

\definecolor{ao(english)}{rgb}{0.0, 0.5, 0.0}
\definecolor{dgreen}{rgb}{0.0, 0.4, 0.05}
\newcommand{\vectr}[1]{\bm{#1}}   
\newcommand{\xten}[1]{\times\!\!10^{#1}}
\newcommand{\ptz}[1]{\left( #1 \right)}
\newcommand{\brkt}[1]{\left[ #1 \right]}
\newcommand\Hmagn{\vectr{H}} 
\newcommand\s{\vectr{\hat{s}}} 

\newcommand\MagnNorm{m} 
\newcommand\Magn{\vectr{\MagnNorm}} 
\newcommand\NeelNorm{l} 
\newcommand\Neel{\vectr{\NeelNorm}} 

\newcommand*\diff{\mathop{}\!\mathrm{d}}
\newcommand{\derivX}[2]{\frac{ \diff #1 }{ \diff #2 } }
\newcommand{\derivt}[1]{ \derivX{#1}{t} }
\newcommand{\derivnX}[3]{\frac{\diff ^{#3} #1 }{\diff{#2}^{#3}}}
\newcommand{\ddt}[1]{ \derivt{#1} }
\newcommand{\ddtt}[1]{ \derivnX{#1}{t}{2} }
\newcommand\hmm[1]{\ifnum\ifhmode\spacefactor\else2000\fi>1000 \uppercase{#1}\else#1\fi}

\newcommand{\NiO}{\ce{NiO}\xspace}
\newcommand{\THz}{\tera\hertz\xspace}
\newcommand{\stt}{\omega_{\tau}}
\newcommand{\mugamma}{\ptz{\mu_0\gamma}}

\newcommand{\thresholdvalue}{{$0.4\xten{30}\mu_B.m^{-2}.s^{-1}$}}
\newcommand{\valueLowestwave}{{$0.11\xten{30}\mu_B.m^{-2}.s^{-1}$}}
\newcommand{\valueHighestwave}{{$8.45\xten{30}\mu_B.m^{-2}.s^{-1}$}}
\newcommand{\chengvalue}{$3.7\xten{30}
\mu_B.m^{-2}.s^{-1}$}

\begin{document}

\title{Ultrafast antiferromagnetic switching in NiO induced by spin transfer torques}

\author{Théophile Chirac}
\email{theophile.chirac@umontpellier.fr}
\altaffiliation{Now at Laboratoire Charles Coulomb, UMR 5221 CNRS-Université de Montpellier, F-34095 Montpellier, France}
\affiliation{CEA Saclay, DRF/IRAMIS/SPEC, UMR 3680 CEA-CNRS, F-91191 Gif-sur-Yvette, France}
\author{Jean-Yves Chauleau}
\email{jean-yves.chauleau@cea.fr}
\affiliation{CEA Saclay, DRF/IRAMIS/SPEC, UMR 3680 CEA-CNRS, F-91191 Gif-sur-Yvette, France}
\author{Pascal Thibaudeau}
\email{pascal.thibaudeau@cea.fr}
\affiliation{CEA--DAM, Le Ripault, BP 16, F-37260, Monts, France}
\author{Olena Gomonay}
\email{ogomonay@uni-mainz.de}
\affiliation{Institut für Physik, Johannes Gutenberg Universität Mainz, D-55099 Mainz, Germany}
\author{Michel Viret}
\email{michel.viret@cea.fr}
\affiliation{CEA Saclay, DRF/IRAMIS/SPEC, UMR 3680 CEA-CNRS, F-91191 Gif-sur-Yvette, France}

\date{\today}

\begin{abstract}
\NiO is a prototypical antiferromagnet with a characteristic resonance
frequency in the THz range.
From atomistic spin dynamics simulations that take into account the
crystallographic structure of NiO, and in particular a magnetic anisotropy
respecting its symmetry, we describe antiferromagnetic switching
at THz frequency by a spin transfer torque mechanism.
Sub-picosecond S-state switching between the six allowed stable spin directions
is found for reasonably achievable spin currents,
like those generated by laser induced ultrafast demagnetization.
A simple procedure for picosecond writing of a six-state memory is described,
thus opening the possibility to speed up current logic of electronic devices
by several orders of magnitude.
\end{abstract}


\maketitle

\section{Introduction}
\label{sec:introduction}

Nature provides us with a variety of magnetic textures, and antiferromagnetism
occurs commonly among transition metal compounds, especially oxides.
It consists in a local combination of magnetic moments of several ions in
crystalline sublattices to produce a vanishing total magnetization.
Such antiferromagnetic (AF) materials display several
interesting characteristics including robustness against external magnetic
perturbations, long coherence times, which make them suitable candidates for quantum computing~\cite{Duong2004,Meier2003}, and picosecond dynamics.
All these advantages make them promising for a new generation of ultrafast
spintronic devices~\cite{Jungwirth2016,MacDonald2011,Gomonay2014,Gomonay2016-06}.
Indeed, thanks to the antiferromagnetic
exchange enhancement~\cite{Gomonay2015,Gomonay2018sep}, the resonance frequency depends both on $\omega_E$ and $\omega_a$ (respectively the exchange
and the anisotropy frequencies, defined from their corresponding energy divided by
the reduced Plank constant $\hbar$). This is to be compared with $\omega_a$ only
for the case of ferromagnets~\cite{Keffer1952,Sievers1963,Cheng2015}.
When $\omega_a\ll\omega_E$, it is proportional to $\sqrt{\omega_E\omega_a}$,
which is generally two orders of magnitude faster than
that for ferromagnets with the same anisotropy frequency.
Therefore, interesting applications can be envisioned from this
dynamical behavior, including building magnetic oscillators in the \THz range
and fast-switching memories~\cite{Cheng2015,Khymyn2017}.
Such devices would be robust against external magnetic fields and compatible
with todays oxide technologies deployed in spintronics.

The past ten years have seen a surge of interest, mainly at a fundamental level,
to bring proofs of concept for using antiferromagnets as memory devices.
Early theories targeted metals~\cite{Nunez2006,Haney2007,Duine2007} and
inspired their validation as memory devices~\cite{Marrows2016,Wadley2016}.
However, insulators may be better candidates
as they exhibit lower magnetization damping and can conduct
spin currents~\cite{Hahn2014,Wang2015,Lebrun2018,Baldrati2019}. 
Many materials are candidates for building memory devices, but so far \NiO has
been the focus of many studies because it is considered as an archetype for
room-temperature applications.
Nevertheless, its full crystallographic form has seldom been considered as far as spintronic
applications are concerned, probably because dealing in detail with the full
magnetic anisotropy landscape can be cumbersome.
Indeed, a single T-domain \NiO is often approximated as an easy plane compound
with a weaker single in-plane easy axis along $[11\overline{2}]$~\cite{Cheng2015,Hutchings1972,Mondal2019}.
It is nonetheless known that this type of domain in \NiO possesses a
sixfold degenerate magnetic state within the easy plane~\cite{Uchida1967}.
This offers a richer switching behavior and also
the possibility to build a six-state memory element
(or at least with three readable states, as $180^\circ$ domains may be hard to distinguish~\cite{Baltz2018}).
The present work aims to harvest these properties by investigating
theoretically the magnetic control of the sixfold symmetry using spin transfer torques.

Experimentally, very recent works have studied the possible influence of a spin injection
on the domain structure of thin \NiO layers.
Spins are usually injected by the spin-orbit torque effect using a Pt layer deposited
on top of the \NiO film.
When a charge current flows in the Pt, the generated transverse spin current induces a non-equilibrium spin
accumulation at the {\NiO}/Pt interface. 
This planar geometry is adequate for the spin Hall effect,
but restrictive in terms of the direction of the injected spins.
Moreover, the required current densities generate a substantial amount of heat
in the structure that may also perturb the AF order. 
We suggest here a different
procedure that relies on the spin injection via ultrafast demagnetization of
an adjacent ferromagnetic (FM) layer by an intense femtosecond laser pulse.
This generates the fastest and strongest spin pulses available so
far~\cite{Kampfrath2010,Kampfrath2013}, with the extra functionality of setting at will
the spin direction in three dimensions (by simply setting the FM magnetization).
Several parameters have to be adjusted in order to optimize the switching mechanism
in the \NiO layer and it is important to identify the most relevant ones,
resulting in both the lowest STT amplitude and the fastest AF switch.
Therefore, the present paper describes the coherent switching processes induced by an
ultrafast laser-generated spin transfer torque in a memory element made of
a bi-layer \NiO/FM.
Our approach relies on numerical atomistic simulations,
where the sixfold symmetry of the \NiO magnetic anisotropy is taken into account.

\section{\NiO crystal structure and magnetic anisotropy}
\label{sec:niocrystal}
\begin{figure}[!ht]
    \begin{tabular}{ll}
    \includegraphics[width=0.49\columnwidth]{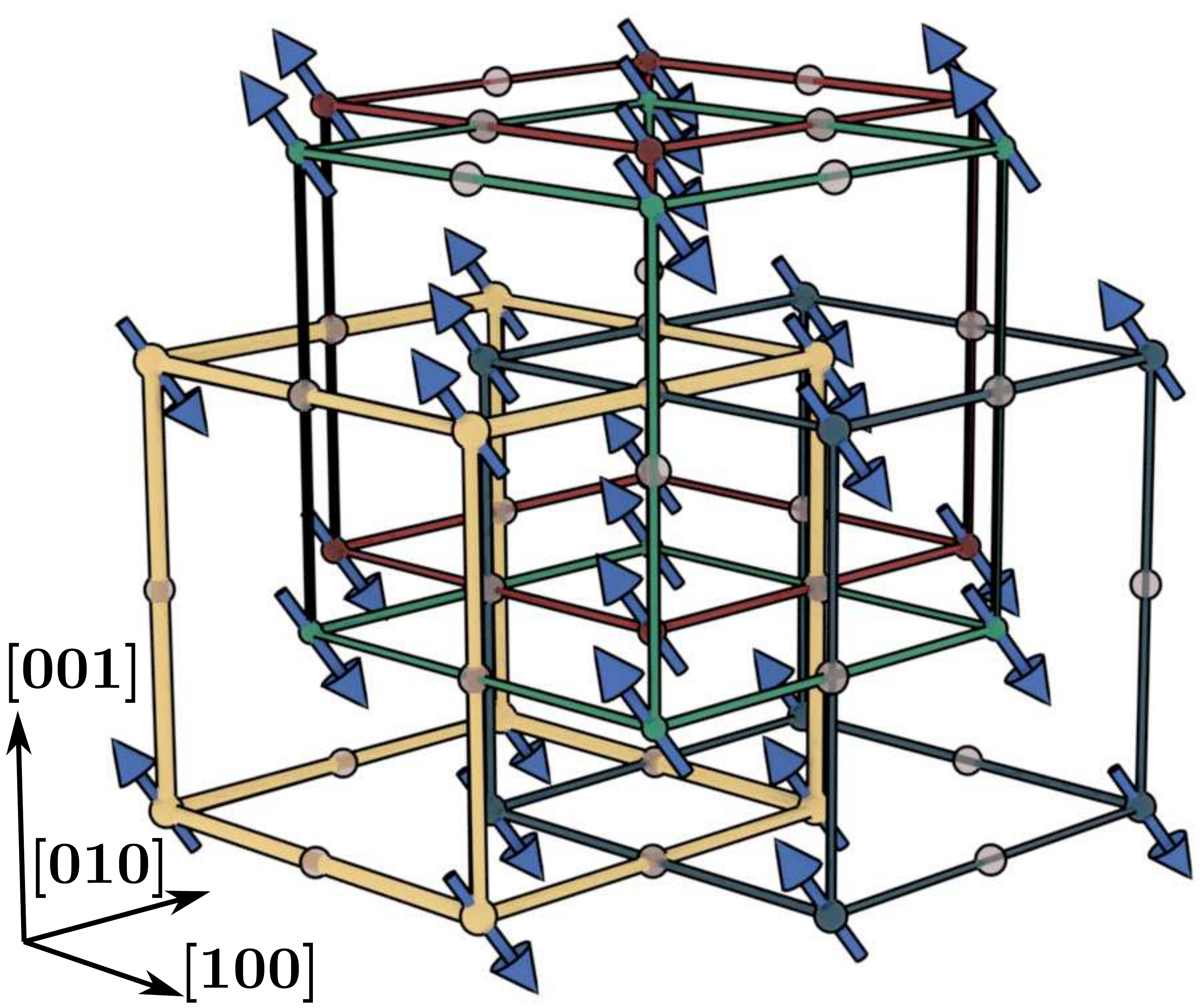}
    &
    \includegraphics[width=0.49\columnwidth]{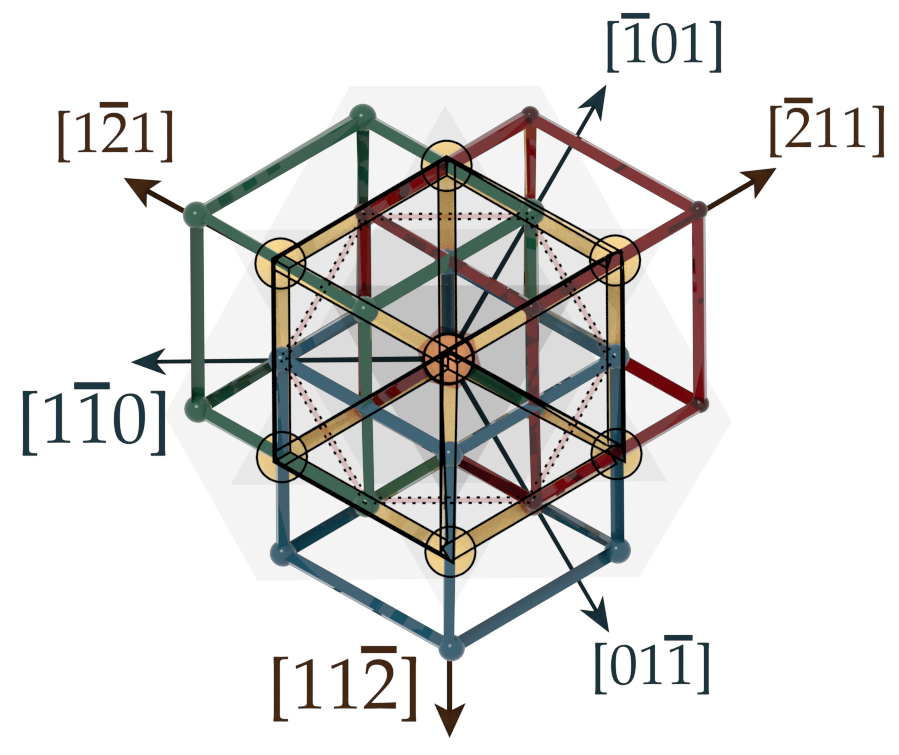}
    \end{tabular}
    \caption{(color online) Crystallographic structure of \NiO. 
    On the left: \NiO has four distinct $180^\circ$ superexchange-coupled
    sublattices (via second nearest neighbors). 
    \label{fig:NiOcubes} 
    On the right: \NiO main crystallographic axes in the $(111)$ plane. The hexagon in dotted lines shows atoms within the same plane. 
    \label{fig:NiOCrystallo} }
\end{figure}

At room temperature, \NiO adopts a fcc structure with $\ce{Ni^{2+}}$ and $\ce{O^{2-}}$ at
the octahedral sites, altered by a slight rhombohedral contraction along one of
the four $[111]$ directions.
This leads to the formation of four possible twin domains (T-domains)
in \NiO crystals~\cite{Hutchings1972}.
In a given T-domain, the magnetic moments of the nickel ions are subject to
various superexchange interactions related to the arrangement of the neighboring oxygen ions.
They consist in a strong antiferromagnetic coupling at $180^\circ$ with
the six {second nearest neighbor (nnn) atoms}, as well as a weak ferromagnetic coupling
at $90^\circ$ with the twelve {nearest neighbors (nn) atoms}, resulting overall in
G-type antiferromagnetism with a staggered order along the $[111]$ direction,
along which ferromagnetic sheets are stacked~\cite{Hutchings1972}.
The associated exchange energies are $J_{nnn}=-19.01\milli\electronvolt$ for the 6 (spin
parallel) next nearest neighbors, $J_{nn}^{-}=1.38\milli\electronvolt$ for the 6
(spin parallel) in-(111)-plane nearest neighbors, and
$J_{nn}^{+}=1.35\milli\electronvolt$ for the 6 (spin antiparallel)
out-of-(111)-plane nearest neighbors~\cite{Hutchings1972}.
The $180^\circ$ nnn-superexchange being by far the strongest, we neglect here
the influence of the nearest neighbor interactions, which is
equivalent to considering only one of the four equivalent sublattices shown
in Fig.~\ref{fig:NiOcubes}.
Even if the nearest neighbor coupling may slightly enrich the
magnetization dynamics, it is considered negligible and is not treated
in the frame of the present paper.

Within one T-domain, \NiO exhibits an anisotropy pattern with a
hard axis along $[111]$, and three easy axes along $[\overline{2}11]$,
$[1\overline{2}1]$ and $[11\overline{2}]$ (right panel of
Fig.~\ref{fig:NiOCrystallo}), defining three possible S-states, and 6 possible
spin orientations.
This configuration is modeled in regard to the $\overline{3}m$ symmetry of
the crystal by taking the expansion of the rhombohedral anisotropy energy to
its leading orders in out-of-plane ($\theta$) and in-plane ($\phi$) components.
Using spherical coordinates in the frame based on
the orthogonal axes $[1\overline{1}0]$, $[11\overline{2}]$ and 
$[111]$, the effective anisotropy energy for a given spin $\vectr{s}_i$
is written as~\cite{Bogdanov1998,Skomski2008} 
\begin{equation}
    E_{K} = -K_{1u}\vectr{s}_i^2\cos^2(\theta) 
    +K_3\vectr{s}_i^6\sin^6(\theta)\cos(6\phi).
    \label{eqn:NiOanisoEn}
\end{equation}

\begin{figure}[!htp]
    \begin{tabular}{c}
    \includegraphics[width=0.3\columnwidth]{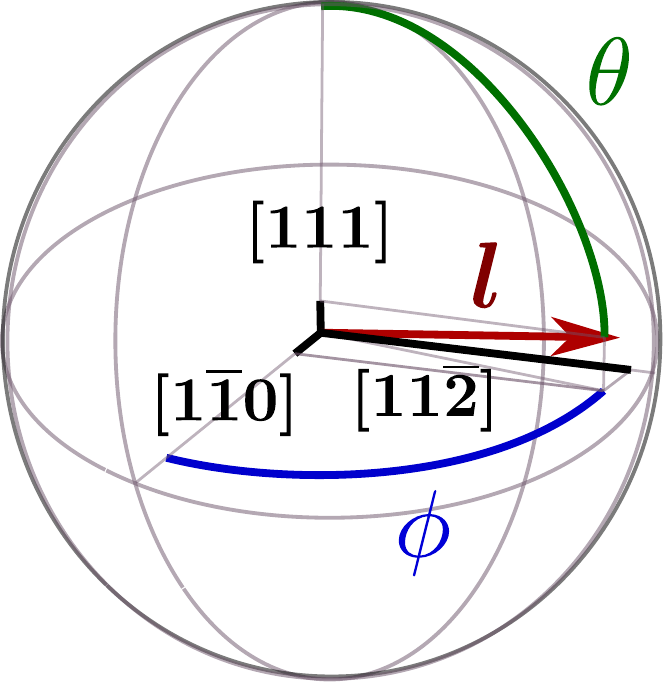}\\
    \includegraphics[width=1\columnwidth]{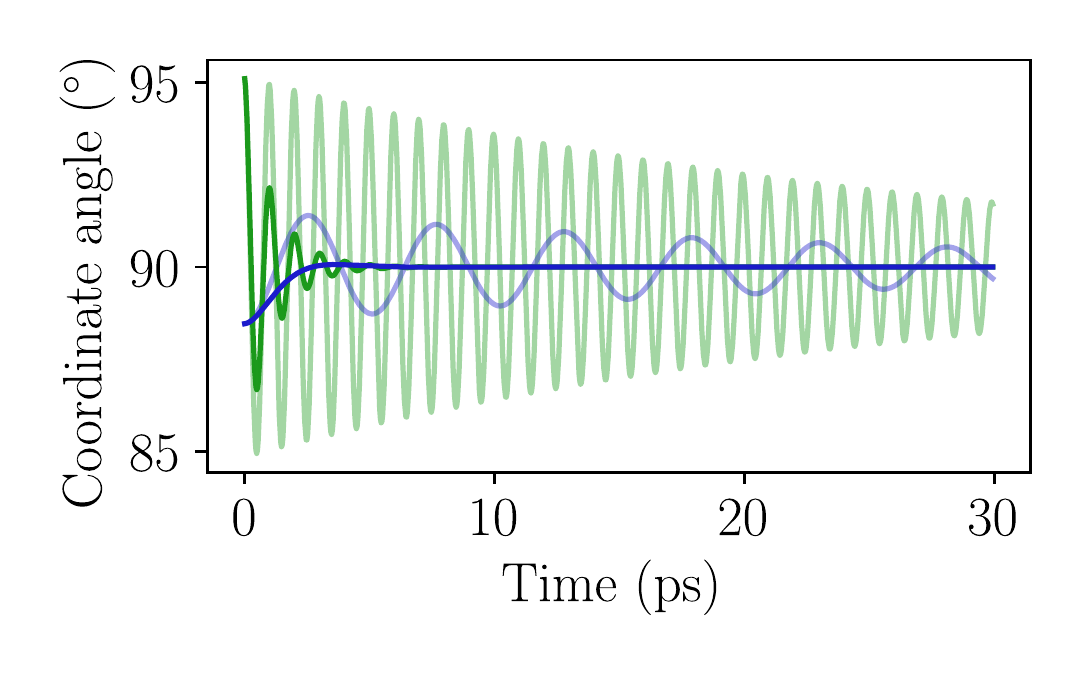}\\
    \includegraphics[width=1\columnwidth]{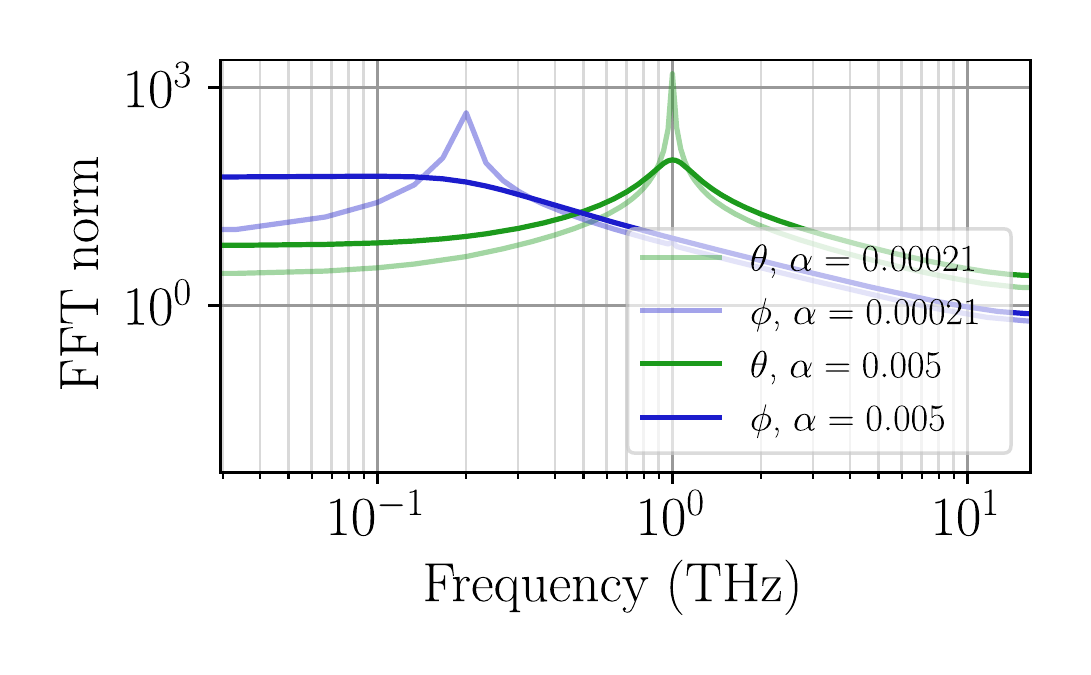}
    \end{tabular}  
    \caption{(color online) Spherical coordinates decomposition of the Néel vector (upper panel). 
        Angular dynamics {\color{ao(english)} $\theta$(t)} and {\color{blue}
	$\phi$(t)}, of the Néel vector of the \NiO antiferromagnetic relaxation,
	starting from a tiny tilt away from equilibrium (middle panel).
	Fourier transform of the angular dynamics, revealing resonances at
	$1\THz$ and $0.2\THz$ at low damping $\alpha=2.1\times 10^{-4}$ (lower panel).
	For practical spintronic applications, the value of $\alpha$, expected around
	$5\times 10^{-3}$, is also computed in the figures.
	A high value for $\alpha$ causes the resonance peaks to flatten and shift.\label{fig:FFT}}
\end{figure}

The values of the anisotropy constants are adjusted dynamically, based on the resonances
observed experimentally for \NiO in
references~\cite{Kampfrath2010,Sievers1963,Satoh2010,Baierl2016,Kohmoto2018,Milano2004,Cheng2015}.
For that purpose, we define the Néel vector
$\vectr{l}\equiv\frac{1}{2}\ptz{\vectr{s_1}-\vectr{s_2}}$ associated to
a set of two spins $\{\vectr{s_1},\vectr{s_2}\}$ representing the two antiferromagnetic
sublattices, and tilt it slightly from its rest position.
A simulation is then performed based on the dynamic model detailed in the next section,
with an effective damping parameter $\alpha=2.1\xten{-4}$ to accommodate
specifically the experimental measurements of Kampfrath et al.~\cite{Kampfrath2010}.
It leads to damped oscillations towards equilibrium with the two expected characteristic
frequencies of $1\THz$ and $0.2\tera\hertz$ when the anisotropy constants are adjusted to
{$K_{1u}=-38\micro\electronvolt$ and $K_3=80\nano\electronvolt$}, as shown in Fig. \ref{fig:FFT}.

With these values, the difference in energy between the $[111]$ and $[11\overline{2}]$
directions is then {$38\micro\electronvolt$} (per atom), and the energy barrier between two stable
neighboring $\langle 11\overline{2}\rangle$ orientations at $60^\circ$ to one another is
{$160\nano\electronvolt$} (per atom). This latter energy is experimentally difficult to measure
because any unrelaxed strain induces a sample dependent larger anisotropy~\cite{Kurosawa1980},
but the former one is of the same order of magnitude as the one found e.g. in
inelastic neutron scattering experiments ($97.2\micro\electronvolt$) ~\cite{Hutchings1972}.
The energy barrier for a coherent switching of a typical AFM containing roughly $10^{5}$ atoms is evaluated to hundreds of kelvin,
which justifies that thermal fluctuations can be neglected in the present simulations.
Based on this description, we will show that magnetic S-states can be dynamically switched under
spin current pulses that are experimentally achievable by ultrafast demagnetization processes
using femtosecond lasers pulses.

\section{Dynamic model}
\label{sec:dynamic_model}

The spin dynamics of antiferromagnets can be described approximately by a set of
two coupled Landau Lifshitz Gilbert (LLG) precession equations linking two
sublattices of equivalent
magnetization~\cite{Chikazumi1997}.
In the case of \NiO, it has been predicted theoretically that a spin current
should produce a spin transfer torque (STT) acting similarly on the two
sublattices and resulting in a significant torque on the Néel vector
$\Neel$~\cite{Gomonay2010,Cheng2014,Khymyn2017}.
In order to tackle the dynamics of this antiferromagnetic order, we consider two
coupled atomistic equations of motion, one for each equivalent magnetic
sublattice labeled by $\s_i$, an unitary vector, that can be formulated as
follows~\cite{Tranchida2018}:
\begin{equation}
\ddt{\s_i} = \vectr{\omega_\text{eff}} \times \s_i \label{eqn:LLG} 
\end{equation}
By denoting $\mu_0$ the vacuum permeability and $\gamma$ the gyromagnetic ratio,
the effective magnetic field on each sublattice is a functional of $\s$, where
$\Hmagn_\text{eff}[\s] =  \vectr{\omega_\text{eff}}[\s_i]/\mugamma$ is
composed of the sum of the anisotropy field
$\vectr{\omega}_{K}/\mugamma$, the exchange field
$\vectr{\omega}_{E}/\mugamma$ and the spin torque, altered by a damping $\alpha$:
\begin{align}
    \vectr{\omega}_{\Sigma} &= \vectr{\omega}_{K}+ \vectr{\omega}_{E} +
    \s_i \times \vectr{\stt}\label{eqn:omegaSigma}\\
    \vectr{\omega}_\text{eff} &= \frac{1}{1+\alpha^2}
    \ptz{\vectr{\omega}_{\Sigma} -\alpha \vectr{\omega}_{\Sigma}\times\s_i}
\end{align}

In detail, each contribution decomposes as follows:

\paragraph{Anisotropy field:} The anisotropy effective field is derived from the functional derivative of eq.\eqref{eqn:NiOanisoEn} with respect to $\s_i$~\cite{Vansteenkiste2014}:
\begin{equation}
\vectr{\omega}_{K}=-\frac{1}{\hbar}\frac{dE_K}{d\s_i}
\end{equation} 

\paragraph{Exchange field:}
The exchange field $\vectr{\omega}_E/\mugamma$ is computed using the Heisenberg
model on the first six neighbors of the superexchange lattice (nnn), with
$J_\text{nnn} = -19.01\milli\electronvolt$~\cite{Hutchings1972,Cheng2015}:
\begin{align}
    \vectr{\omega}_E = \frac{J_\text{nnn}}{\hbar}\sum_{j=1}^6{\s_{j}} 
\end{align}

\paragraph{Spin torque:}
$\vectr{\stt}$ represents the frequency in the Slonczewski's spin transfer
torque expression~\cite{Slonczewski1996,Gomonay2010}.
For a STT $\vectr{\omega}_s$ (expressed in $\mu_B.m^{-2}.s^{-1}$) injected though
a thin layer of \NiO from an adjacent ferromagnetic layer, we can estimate it as:
\begin{equation}
\vectr{\stt} \simeq \frac{G}{d} \frac{a^3}{n_s}\vectr{j}_s \label{eqn:STT}
\end{equation}
where $G$ is the spin transparency of the interface, $a$ the lattice constant,
$n_s$ the number of magnetic atoms per unit cell, $d$ the layer thickness and vector $\vectr{j}_s$ is parallel to the spin current polarization, with a magnitude equal to the spin current density. In the present paper, values are expressed directly in spin currents
taking $a = 4.177 \angstrom$, $n_s = 4$,
$d = 2 \nano\meter$ and $G = 0.1 \mu_B^{-1}$.
The \NiO thickness is optimally taken close to the experimentally estimated
penetration depth of spin-polarized electrons~\cite{Hahn2014,Wang2015}.

For all the following simulations, which involve thin films,
the damping value is set to $\alpha=0.005$. 
This value is higher than the one used to adjust the resonances, which corresponded
to a value typically found in bulk samples.
With this higher value, we also expect to account for several additional
mechanisms, including for example the spin dissipation induced by an adjacent ferromagnetic layer.
This value appears sufficient to capture a broad range of possible effects
encountered in thin films spintronics (even though we recognize that the Gilbert form here adopted
is not quite proper to accurately account for inter-lattice dissipations~\cite{Kamra2018}).

\section{Results and discussion}
\label{sec:results}

\begin{figure}[!ht]
\begin{tabular}{c}
\includegraphics[width=\columnwidth]{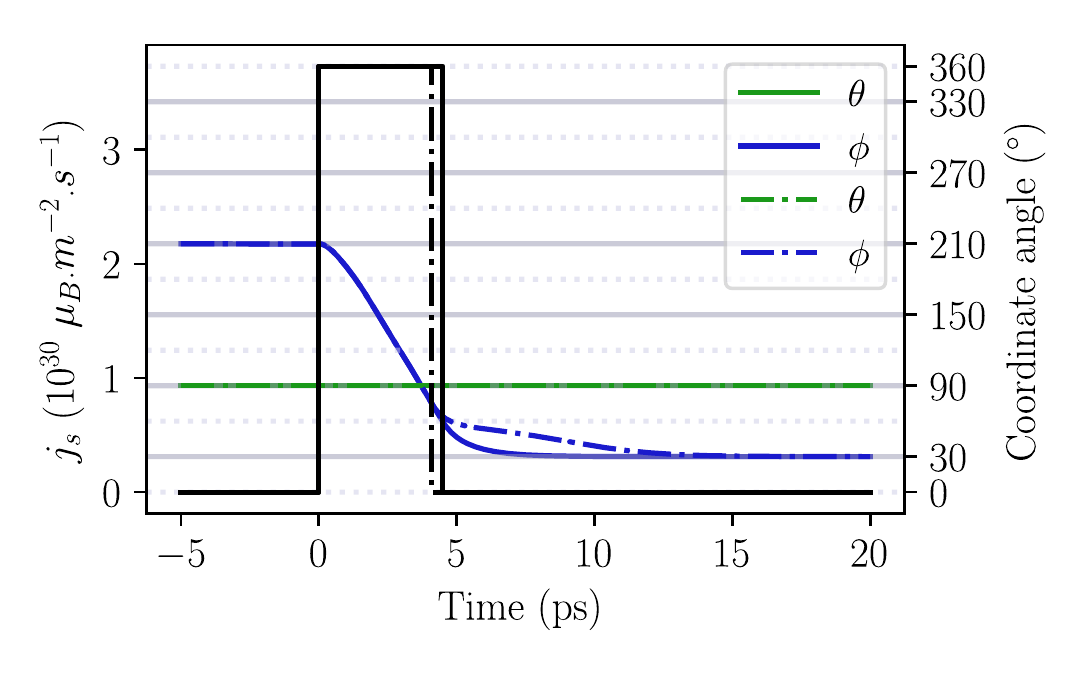}\\
\includegraphics[width=\columnwidth]{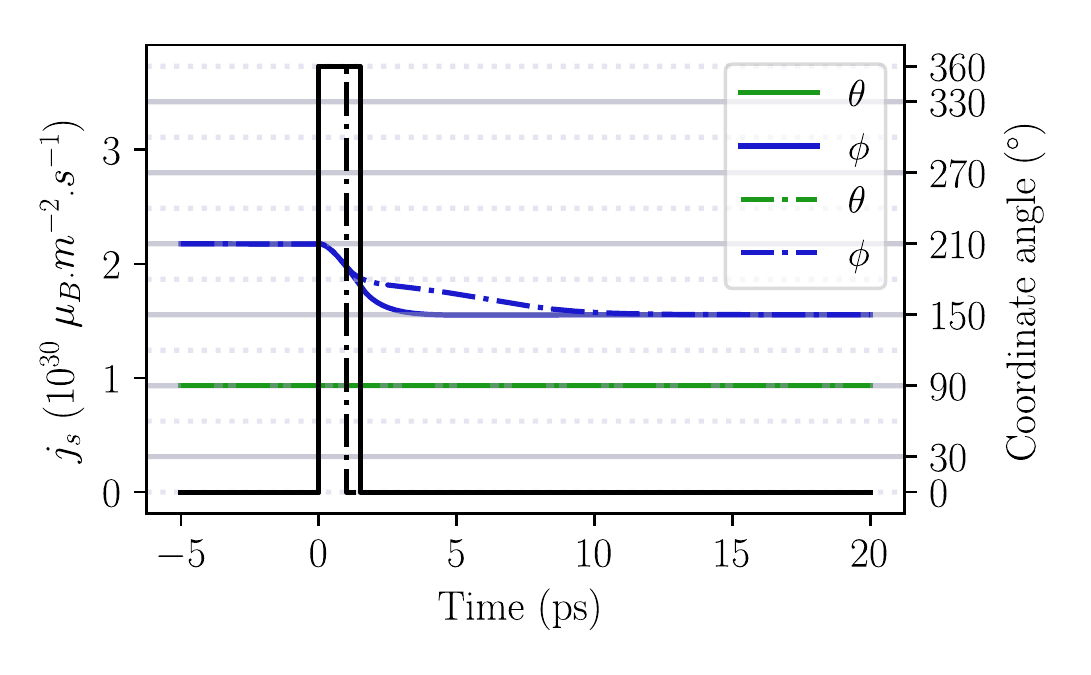}
\end{tabular}
\caption{(color online) \NiO switching with a $\vectr{j}_s$ along $[111]$ for $j_s=${\chengvalue}. 
Black curves show the switching on and off of the STT. 
A duration of {$4.5ps$} gives a $180^\circ$ switch (upper panel),
whereas a {$1.5ps$} pulse is enough to trigger a $60^\circ$ switch (lower panel).
The same final $\theta$ states can be reached when the pulse durations are reduced to 
{$4.1\pico\second$} and {$1\pico\second$} respectively, at the cost of a longer relaxation time.}
\label{fig:NiOpulsedSwitch}
\end{figure}

Within this dynamic model for \NiO, a STT $\vectr{\omega}_s$ applied along the
$[111]$ direction of a T-domain can trigger a change of orientation of the spins,
switching from one S-state to another. 
This is the case studied analytically by Cheng et al.~\cite{Cheng2015} albeit in an orthorhombic symmetry. 
Our anisotropy profile exhibits the 6 possible stable $\langle
11\overline{2}\rangle$ orientations, and a switch between them
can be achieved in a picosecond timescale, as revealed by the Fig.~\ref{fig:NiOpulsedSwitch}.

Due to the presence of intermediate stable positions,
the minimum duration of STT needed to achieve a $180^\circ$ switch is significantly reduced
compared to the one predicted in ref.~\cite{Cheng2015}.
For the same spin current value of {$j_s=$\chengvalue} considered in this reference,
the minimum duration is reevaluated from $10\pico\second$ to {$4.5\pico\second$}.
Even shorter switches can be achieved when reorienting the spins by $60^\circ$.
In this case, the duration of the STT pulse can be reduced even to $1\pico\second$,
with the same intensity.

\begin{figure}[!ht]
    \includegraphics[width=\columnwidth]{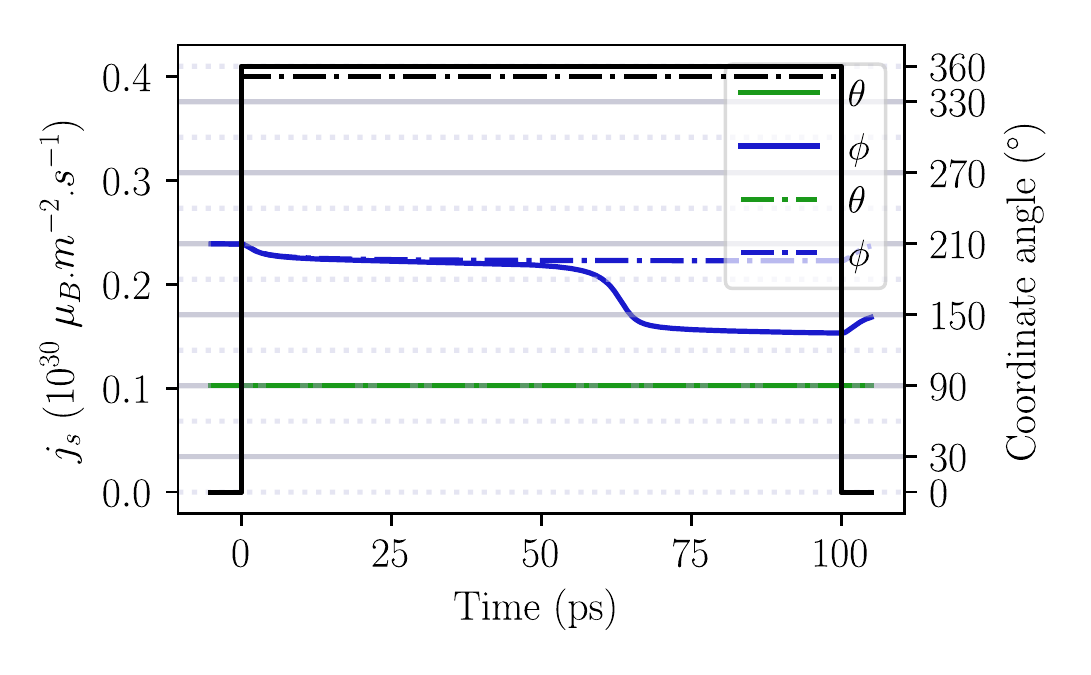}
    \caption{(color online) A $\vectr{j}_s$ of {$0.40\xten{30} \mu_B.m^{-2}.s^{-1}$} is below 
    the threshold value to initiate a switch (dotted lines), whereas
    {$0.41\xten{30} \mu_B.m^{-2}.s^{-1}$} is above the threshold (full
    lines). 
    }
    \label{fig:NiOLimitSTT}
\end{figure}

As the threshold for switching is directly linked to the anisotropy value,
the lowest STT amplitude is obtained when the \NiO spin trajectories remain in the easy plane.
This is indeed achieved when the spin current is polarized along the $[111]$ direction and
for a threshold close to \thresholdvalue, as shown in Fig.~\ref{fig:NiOLimitSTT}.
As long as the STT excitation exceeds the threshold, precession occurs at a frequency depending on
how much the system is driven above the threshold, as well as its natural timescale and damping.
Once the spin pumping is turned off, the system precesses permanently for zero damping, whereas it falls quickly to an equilibrium position for large damping. For the realistic value of $\alpha\approx 0.005$ and by providing a suitable spin pulse strength and duration, all the in-plane equilibrium angles can be reached at will in some picoseconds.
Interestingly, it is in principle possible to apply a bipolar spin current pulses in order to fall more reliably into the chosen position.

\begin{figure}[!ht]
	\includegraphics[width=\columnwidth]{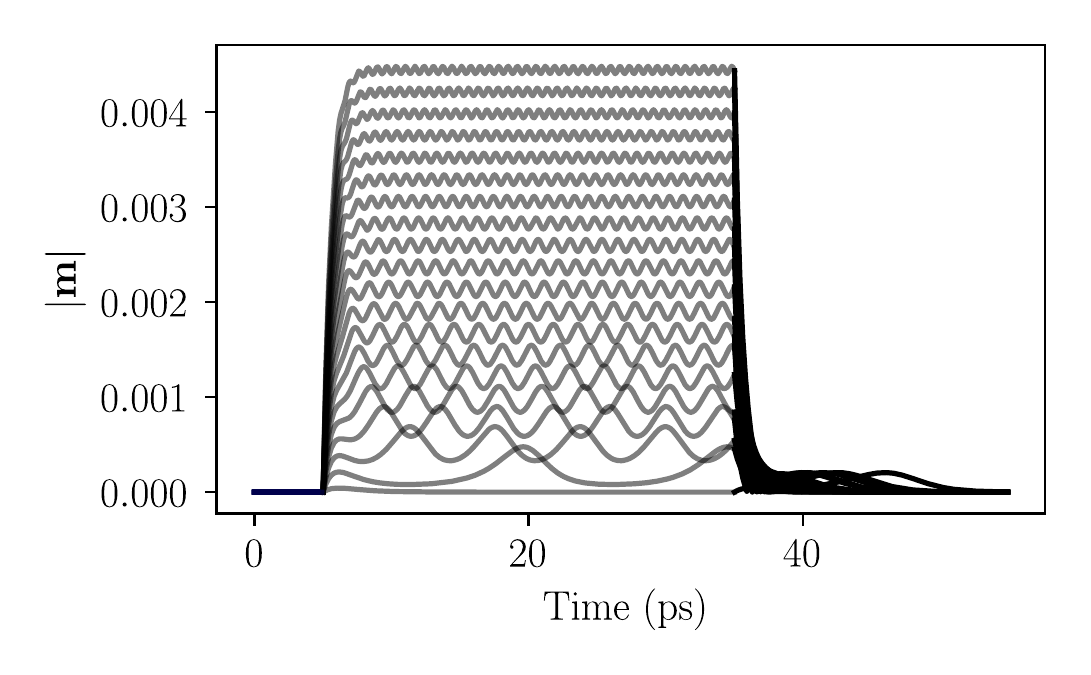}
	\caption{(color online) Evolution of the norm of the average magnetization vector $\Magn=\frac{1}{2}\ptz{\s_1+\s_2}$ for different $30\pico\second$-long pulses, with $j_s$ varying from {\valueLowestwave to \valueHighestwave}.
All the rise and decay stages match an exponential law with an identical time constant of $0.6\pico\second$.}
\label{fig:MnormForJ}
\end{figure}

Some simple expectations can also be inferred directly from the differential equations of motion of the angular dependence of the Néel vector,
as shown in appendix \ref{sec:RaisingTime}.
Firstly, as far as writing speed is targeted, one may realize that for STT
pulses sufficiently fast not to lose too much angular momentum
in damping processes, i.e. much faster than $1/\ptz{2\alpha \omega_{E}}$, only
the total number of injected spins matters.
Indeed, in that case the STT cants the two sublattices with a characteristic time of $1/(2\alpha\omega_E) = 0.6\pico\second$, as
shown in Fig.\ref{fig:MnormForJ}.
This stores in the system's magnetization a quantity of exchange energy 
proportional to the number of
injected spins.
Once the driving is turned off, this energy drives the precessional motion of the Néel vector
at its natural precession frequency
${\sim\propto\sqrt{\omega_{a}\omega_{E}}}$, until the damping fully stops the precession.
This dynamics is quite similar to what was predicted for
noncollinear antiferromagnets~\cite{Gomonay2015,Kimel2009}.

\begin{figure}[!ht]
    \includegraphics[width=\columnwidth]{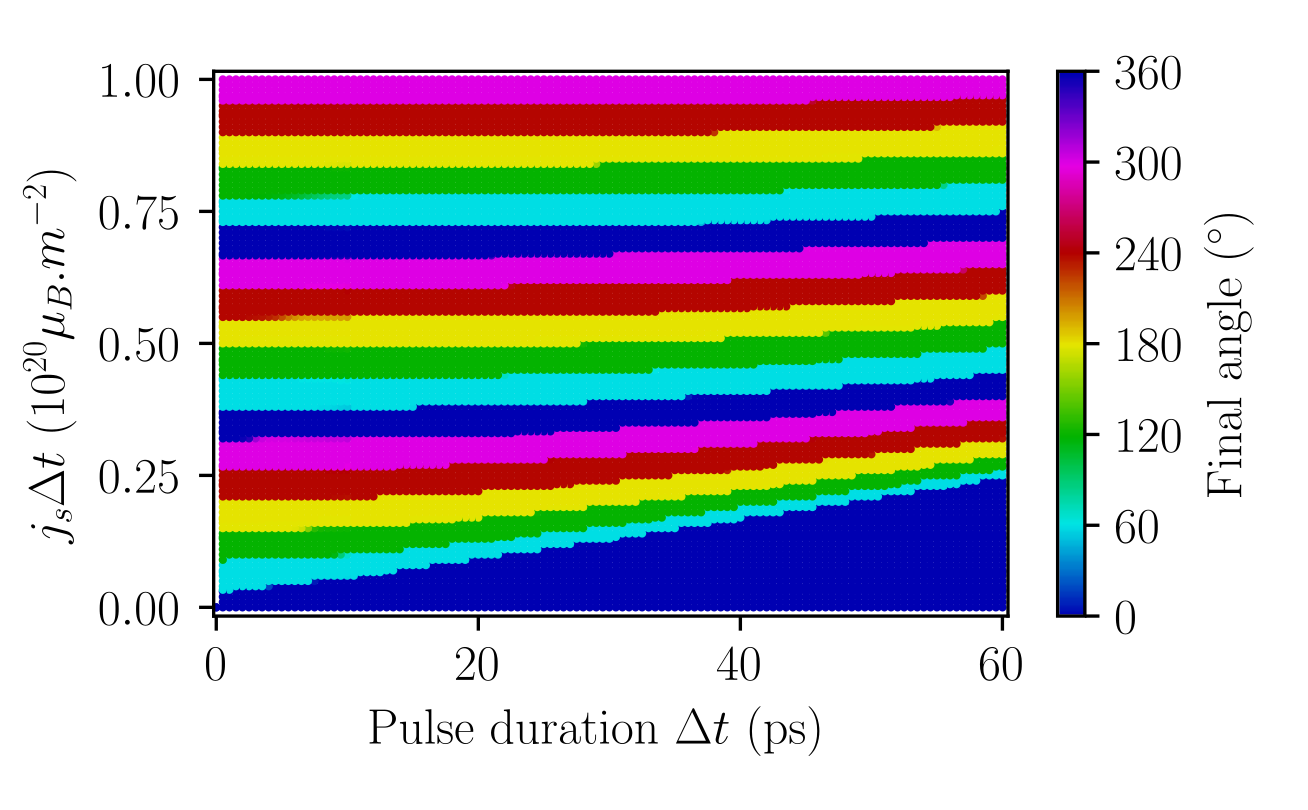}
    \caption{(color online) Final orientation $\Delta\phi$ with respect to the number of injected spins $j_s\Delta t$ and the duration $\Delta t$ of the injection. Above the \thresholdvalue threshold value, a constant number of injected spin gives approximately a constant switch.}
    \label{fig:PhaseDiagramNbSpin}
\end{figure}

\begin{figure*}[!htbp]
    \includegraphics[width=\textwidth]{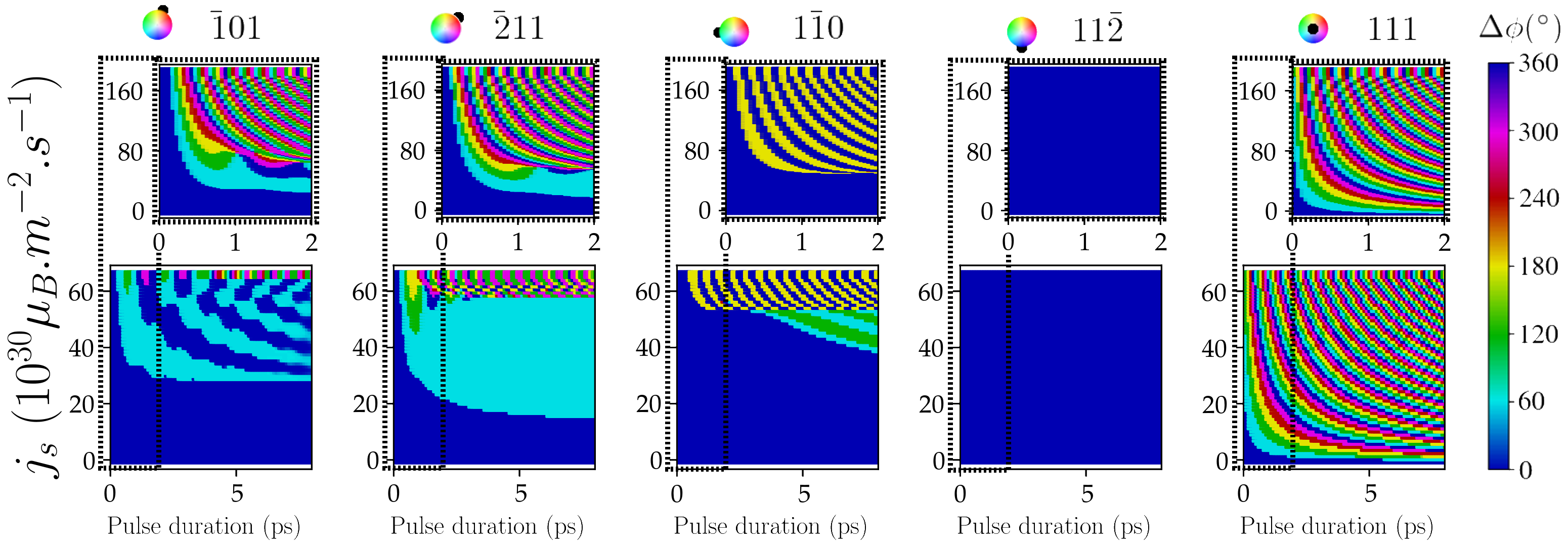}
    \caption{(color online) S-state switch phase diagram for sub-picosecond gate pulses of spin current in the main in-plane angles.} 
    \label{fig:RelaxPhaseDiagAngle}
\end{figure*}

The horizontal lines on Fig.~\ref{fig:PhaseDiagramNbSpin} show that the requirement to reach a given memory state, depends only on the total number of injected spins $j_s\Delta t$, for $j_s$ far above the \thresholdvalue threshold value (for a $2\nano\meter$ thick \NiO).
One can conclude that for pulses faster than a few picoseconds, no pulse shaping is necessary and the only parameter governing the switching is the total number of injected spins. Therefore, the injection can be achieved in an arbitrarily short period of time: the shorter the pulse duration is, the stronger the STT strength must be, as shown in Fig.~\ref{fig:RelaxPhaseDiagAngle}. 
After the injection, the dynamics proceeds, until all the accumulated STT energy stored in the canting is damped, on a timescale determined by $\alpha$. Consequently, a bit of information can take less than a fraction of
$2\pi\sqrt{\omega_{a}\omega_{E}}$ picoseconds to
reach a new value, depending on how far from equilibrium the STT ends.
Nonetheless, the final rest time to reach a stable state is incompressible and
depends on the damping value.
As far as stabilization speed is concerned, a too low damping is therefore not desirable,
and a value higher than 0.005 should be optimal~\cite{Cheng2015}.
One could then envision to write a logical bit very fast, but a few picoseconds
waiting time must be observed before the bit acquires stability.
As the total rest time is set by the damping, it is not possible to shorten the total switching procedure.
Another option to improve fast switching would be to use tailored shaped bipolar pulses to reduce quickly to zero the inertia stored in the spin canting and force the system to reach an equilibrium minimizing the ringing. This subpicosecond fine tuning, however, seems presently out of reach experimentally.
 
Moreover, fully deterministic switching is a particularly difficult problem~\cite{Chen2018}.
This stems from the absence of the internal self-stabilization mechanism present
in ferromagnets ~\cite{Chen2018}. In this respect, it is instructive to consider
other directions for the STT to force the AF vector to take a trajectory through
higher anisotropy energies, as shown in Fig.~\ref{fig:RelaxPhaseDiagAngle}.
There, the final states for a $\vectr{\stt}$ along one of the main in-plane axes are displayed.
For directions other than $[111]$, the threshold values are much higher and often experimentally
out of reach. Especially when the STT is applied parallel to
the spins direction ($[11\overline{2}]$), the excited mode generally generates
a cone of precession much smaller than $60^\circ$, which does not lead to switching.
For the other directions, the spins tend to precess around the STT, but with trajectories
constrained by the anisotropy profile. Precessing out of the easy plane requires more energy,
as can be seen in Fig.~\ref{fig:RelaxPhaseDiagAngle} for the $[1\overline{1}0]$ direction.
STT directions at $30^\circ$ or $60^\circ$ to the spin are more efficient.
Indeed, they generate a sufficiently small precession cone to remain close to the easy plane.
When at $30^\circ$ (direction $[\overline{1}01]$), the spins can easily oscillate between
the two neighboring positions.
Finally, the direction at $60^\circ$ (direction $[\overline{2}11]$) is particularly
interesting for controlled writing application.
There, the STT causes a sufficiently large precession to induce a switch,
with a trajectory experiencing a reduced torque as it gets close to the STT axis.
This enhances a more efficient trapping from the stable state along the STT,
as visible on the corresponding diagram of Fig.~\ref{fig:RelaxPhaseDiagAngle}. 

\begin{figure}[htb]
    \includegraphics[width=\columnwidth]{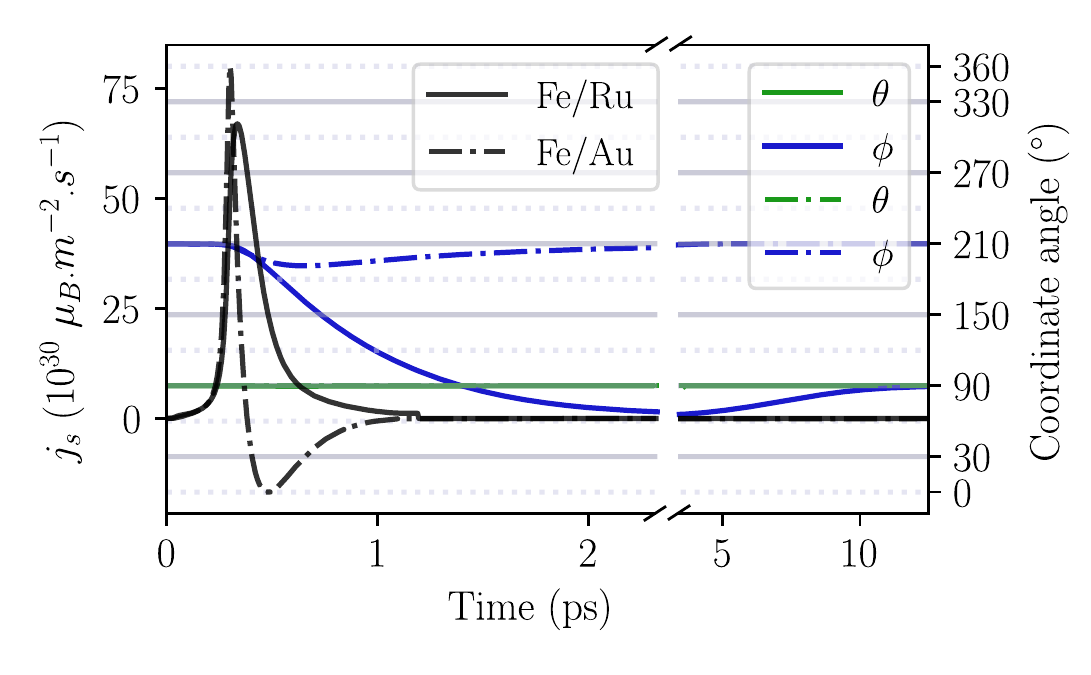}
    \caption{(color online) AF states switching mechanism after excitation profiles inspired
    from those computed
    in \ce{Fe/Ru} and \ce{Fe/Au} by reference~\cite{Kampfrath2013} (see text).}
    \label{fig:KampfrathSwitches}
\end{figure}

In the light of the present simulations, it is important to assess whether or not
the conditions for writing such a memory could be achieved experimentally.
The shortest spin transfer torque stimulus experimentally available is that generated by
the ultrafast demagnetization of a ferromagnetic layer by a
femtosecond laser pulse~\cite{Kampfrath2013}. Emitted from the ferromagnetic layers, bursts of spins have been injected
into different metals using double layers (e.g. \ce{Fe/Ru} or \ce{Fe/Au}), where their spin conversion generates a \THz pulse of electric charge.
Hence, the heavy metal layer acts as a sensor for the spin current burst. Using the two reported shapes, we run our simulations in order to estimate if this technique can be adequate for addressing a memory element made of NiO.
The results, displayed in Fig.~\ref{fig:KampfrathSwitches}, indicate that
the unipolar spin burst generated in a \ce{Fe/Ru} structure applied in
the $[111]$ direction of \NiO, can effectively switch the Néel vector $\Neel$ to
another stable position. On the other hand, the bipolar pulse of
the \ce{Fe/Au} structure cannot.
 This is consistent with our previous observation that for such short pulses, only the total amount of injected spins is relevant. 
For the bipolar pulse, this quantity is too small. 

This is therefore an encouraging result, although a real spin current shape cannot be directly inferred from those observed in metallic double layers. A more realistic \ce{CoFeB}/\NiO system should be tested as the spin injection efficiency should be reduced because of a poorer interface transparency.
Nevertheless, as the minimum number of injected spins for switching is four times below that of the experimental spin bursts in \ce{Fe/Ru}, our simulations indicate that very fast switching should be possible in \NiO, when an adjacent
ferromagnetic layer is subjected to ultrafast demagnetization.

\begin{figure}[ht]
    \includegraphics[width=\columnwidth]{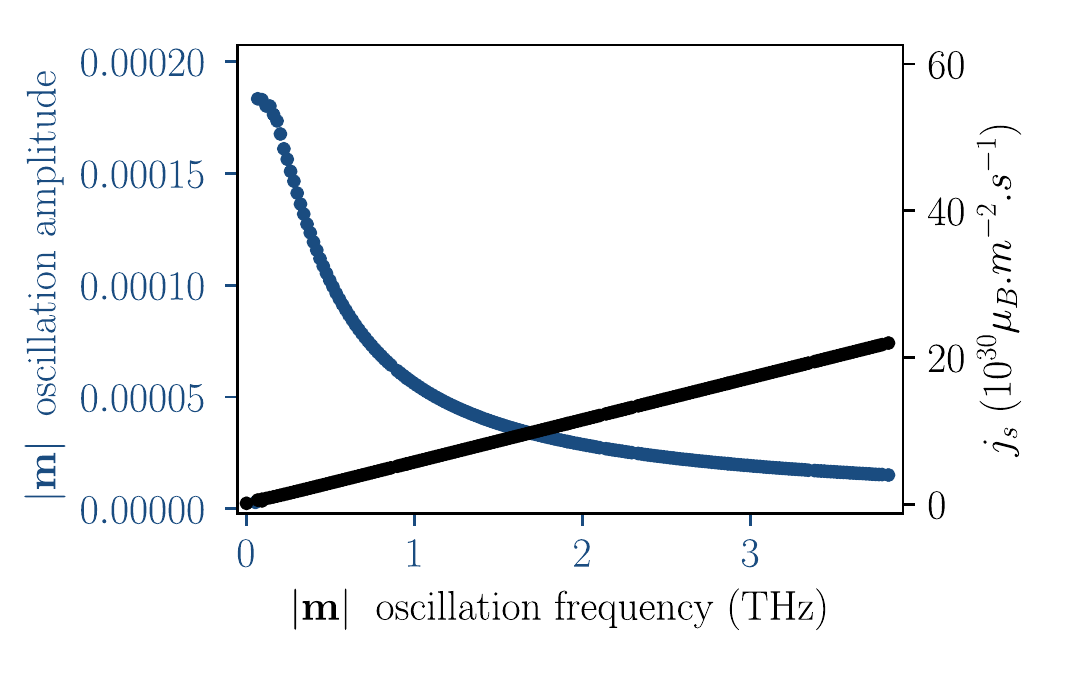}
    \caption{(color online) Amplitude and frequency of the oscillations of $|\Magn|$ for
    different spin currents.}
    \label{fig:BumpResonnance}
\end{figure}

\begin{figure}[ht]
    \includegraphics[width=\columnwidth]{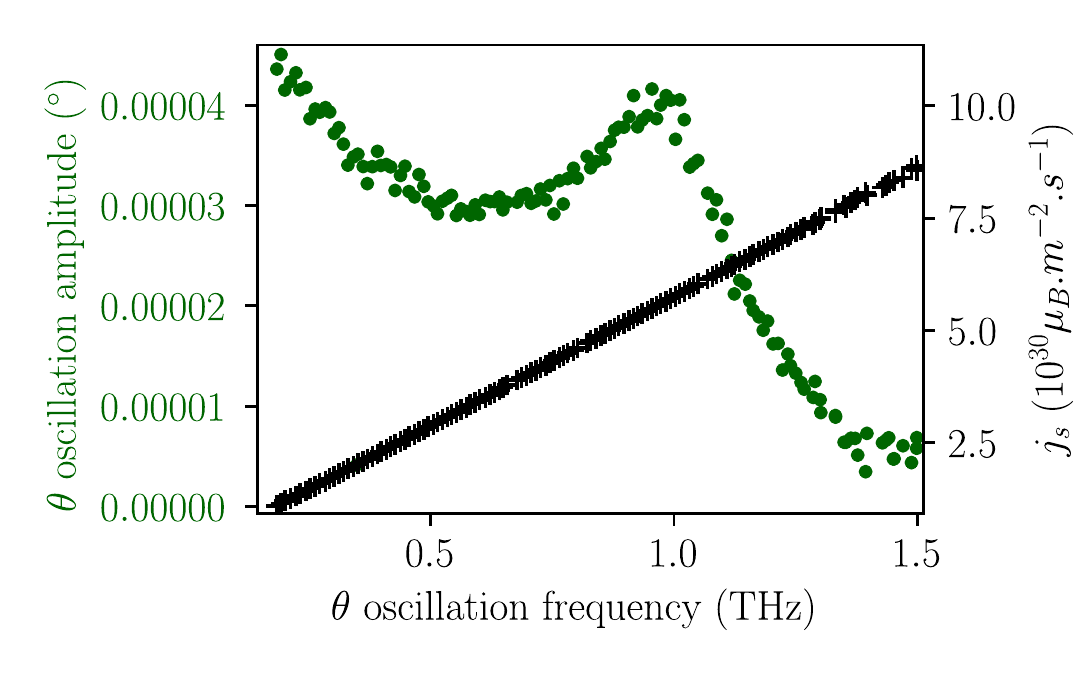}
    \caption{(color online) Amplitude and frequency of the out-of-plane oscillations for
    different spin currents. }
    \label{fig:MResonnance}
\end{figure}

Finally, similar systems can also be used for $\THz$ oscillators, as reported in reference~\cite{Khymyn2017}. In that case, the characteristic setting time $1/(2\alpha\omega_E)=0.6\pico\second$ must be taken into account before observing stable oscillations. Fig.\ref{fig:MnormForJ} and \ref{fig:BumpResonnance} show the behavior of the uncompensated magnetization when the \NiO is pumped with long duration pulses. The frequency of the oscillations varies linearly with the spin current intensity and can be hypothetically adjusted at will.  Nonetheless, the amplitude of the oscillations tends to be higher for low spin current intensities. For the low currents just above the threshold, $|\Magn|$ spikes periodically with high amplitude.
For these values, $\phi$ indeed undergoes rapid accelerations when passing $\langle 11\overline{2}\rangle$, but the pace is low since it is slowed down every time it passes the anisotropy barriers near $\langle 1\overline{1}0\rangle$.
Hence, the duty cycle is reduced and the harmonicity is degraded. This can be seen as a periodical pulses generation.
As shown on Fig.~\ref{fig:MResonnance}, the mode at $1\THz$ is excited by the out-of plane excursion of $\Neel$ during its in-plane rotation.

\section*{Conclusion}
\label{sec:conclusion}

By performing atomistic spin simulations, we have shown that a thin layer
of \NiO can in principle be used to build a six-state memory device.
By using magnetic anisotropy expressions that reflect the real symmetries of
the material, we have exhibited that experimentally available sub-picosecond pulses are
\textit{a priori} adequate to switch a $2\nano\meter$ thick memory element.
Thus, we propose a device formed by a \NiO/ferromagnetic double layer,
where an ultrafast laser is used to inject a spin population at an arbitrary spin angle,
by demagnetizing the ferromagnetic layer. 
Both constraints on the growth of epitaxial \NiO, as well as on the control of
the STT direction are then released by this technique.
The excitation process offers the possibility to access deterministically the
six AF spin states at picoseconds time scale.
Beyond memory devices, the non-trivial magnetic anisotropy of \NiO suggests
a richer dynamics that could lead to other spintronic applications in the \THz range.

\begin{acknowledgments}
We wish to acknowledge Julien Tranchida for the fruitful discussions, and the French National Research Agency for support with the project ANR SANTA (Grant No.\href{https://anr.fr/Projet-ANR-18-CE24-0018}{ANR-18-CE24-0018-03}). 
O.G. acknowledges the Alexander von Humboldt Foundation,
the ERC Synergy Grant SC2 (No.\href{https://cordis.europa.eu/project/id/610115}{610115}) and funding by the Deutsche Forschungsgemeinschaft (DFG, German Research Foundation) - \href{https://www.dfg.de/en/funded_projects/current_projects_programmes/list/projectdetails/index.jsp?id=268565370}{TRR 173 - 268565370} (project \href{https://www.uni-kl.de/trr173/research/projects/project-b12-2020-2023/}{B12}).
\end{acknowledgments}

\appendix

\section{Numerical implementation}
\label{sec:numerical_implementation}

\begin{figure}[htb!]
\centering
\begin{tabular}{c}
\includegraphics[width=\columnwidth]{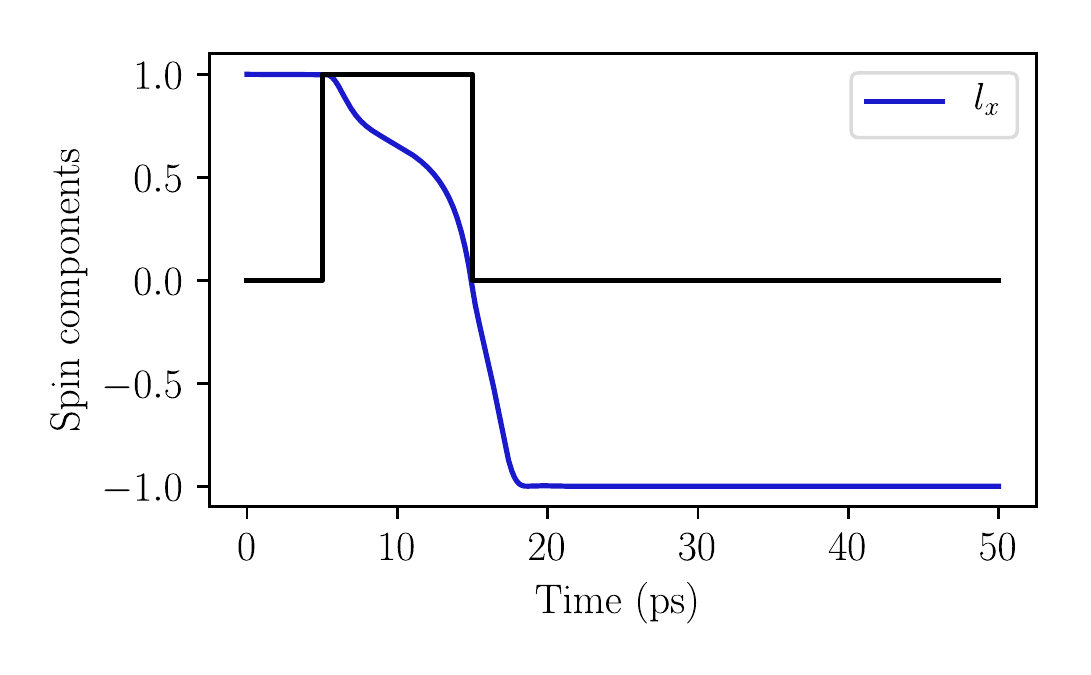}\\
\includegraphics[width=\columnwidth]{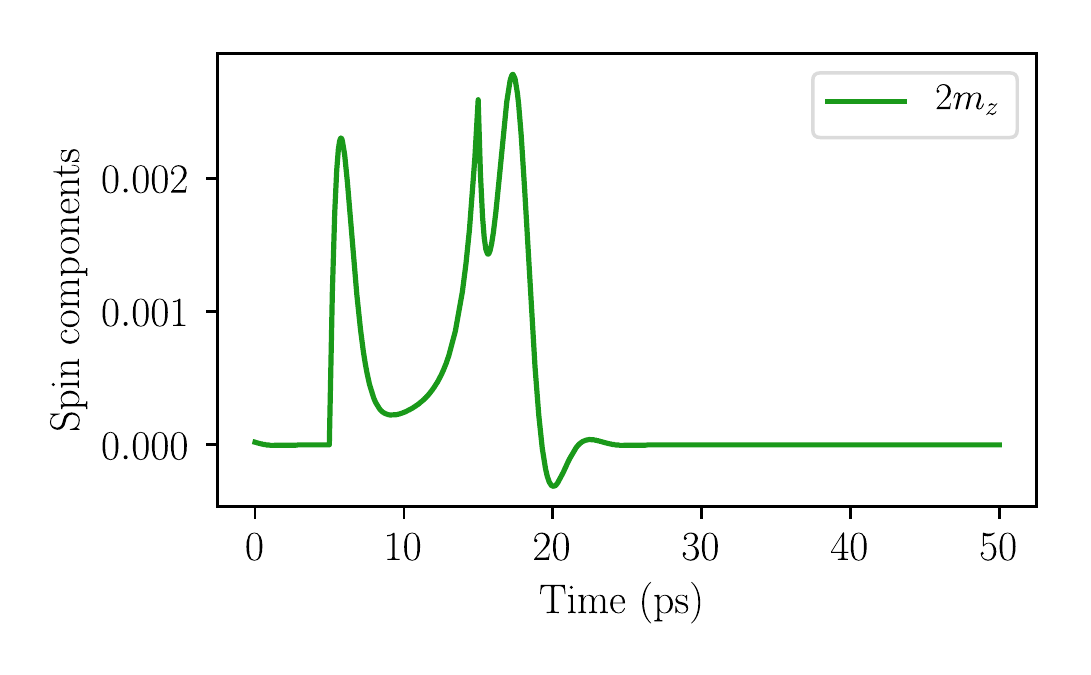}
\end{tabular}
\caption{(color online) Reproduction of the switching process described in reference
\cite{Cheng2015}.
The upper panel displays the STT pulse (in black), the x-component of the Néel
vector ${\bm l}$ (in blue).
The lower panel displays the z-component of twice the average AF magnetization ${\bm m}$.}
\label{fig:RanCheng} 
\end{figure}

Simulations are performed for two spins that are coupled with effective fields.
Each spin represents its own ferromagnetic sublattice.
The equations of precession are integrated in time with a symplectic integrator.
The transverse equation (\ref{eqn:LLG}) is discretized to update only the
orientation of each spin for a given timestep $\Delta t$.
In practice, $\vectr{s}_{t+\Delta t}$ is computed from $\vectr{s}_{t}$ and
$\vectr{\omega_{\mathbf{eff}}}$ with $\mathcal{O}({\Delta t}^3)$ precision
\cite{Tranchida2018}, according to:

\begin{equation}
\begin{split}
\vectr{s}_{t+\Delta t} = 
&\frac{1}{1+\frac{1}{4}\ptz{\Delta t}^2\vectr{\omega_{\mathbf{eff}}}^2}\left[\vectr{s}_{t}+
\Delta t \ptz{\vectr{\omega_{\mathbf{eff}}}\times\vectr{s}_t} \phantom{\frac{1}{1}}\right.\\
&\left.+\frac{1}{4}\ptz{\Delta t}^2\ptz{2\ptz{\vectr{\omega_{\mathbf{eff}}}\cdot\vectr{s}_t}
\vectr{\omega_{\mathbf{eff}}}-\vectr{\omega_{\mathbf{eff}}}^2\vectr{s}_t }\right]
\end{split}
\end{equation} 
To check the consistence of this approach, we evaluate the dynamics of the Néel
vector and average magnetization by using the numerical values found in
reference~\cite{Cheng2015,Nussle2019}.  
Our simulations reproduce well the published results as shown in
Fig.~\ref{fig:RanCheng}.

The simulations in the core of the paper were done with a time step of $1\xten{-16}\second$, on a total of two atoms only, with an exchange value accounted 6 times, which is equivalent to periodic boundary conditions in all directions, for the given superexchange sublattice.

\section{Raising time in an uniaxial anisotropy}
\label{sec:RaisingTime}
By considering a {sixfold} easy axis ($\omega_a$ along $x$) in a hard plane
($\omega_A$ along $z$), as long as $|\Magn|\ll|\Neel|$, the equation for $\Neel$
reduces to a one dimensional problem~\cite{Cheng2015,Gomonay2018sep,Yamane2019}:
\begin{equation}
\ddtt{\phi}+2\alpha\omega_E\ddt{\phi}+\frac{\omega_R^2}{2}\sin{\ptz{{6}\phi}} 
= 2\omega_E\omega_\tau,
\end{equation}
with $\omega_R\equiv\sqrt{2\omega_a\omega_E}$.
The out-of-plane component of the magnetization vector is simply
$\MagnNorm_z=-\ptz{2\omega_E+{|\omega_A|}+\omega_a l_x^2}^{-1}\ddt{\phi}$.
With $\phi(0)=0$ and by considering the response near the beginning of the
pulse, for which $\phi\ll 2\pi$, the previous differential equation is
linearized, so that:
\begin{equation}
\ddtt{\phi}+2\alpha\omega_E\ddt{\phi}+{3}{\omega_R^2}\phi = 2\omega_E\omega_\tau,
\end{equation}
and solved, after defining
$\omega_{ip}=\sqrt{{6}\omega_a\omega_E-\alpha^2\omega_E^2}$.
We find
\begin{equation}
\phi(t) = \frac{\omega_\tau}{\omega_a}\ptz{1-e^{-\alpha\omega_E t}
\brkt{\cos{\ptz{\omega_{ip}t}}+\frac{\alpha \omega_E}{{\omega_{ip}}}
\sin{\ptz{\omega_{ip}t}}}}
\end{equation}
Therefore near $t=0$,
\begin{equation}
\ddt{\phi} \sim 2\omega_\tau\omega_E t,\label{eqn:raisingTime}
\end{equation}
meaning that from reference~\cite{Khymyn2017}, a simple model for
the convergence to the average value of the angular velocity
$\frac{\omega_\tau}{\alpha}\ptz{1-e^{-t/\tau_c}}$ gives
$\tau_c=1/(2\alpha\omega_E)\sim 0.6\pico\second$, which is in agreement with our
numerical simulations, as depicted in Fig.~\ref{fig:MnormForJ} in section
\ref{sec:results}.

\bibliographystyle{apsrev4-1}
\bibliography{paper}  

\begin{thebibliography}{45}%
\makeatletter
\providecommand \@ifxundefined [1]{%
 \@ifx{#1\undefined}
}%
\providecommand \@ifnum [1]{%
 \ifnum #1\expandafter \@firstoftwo
 \else \expandafter \@secondoftwo
 \fi
}%
\providecommand \@ifx [1]{%
 \ifx #1\expandafter \@firstoftwo
 \else \expandafter \@secondoftwo
 \fi
}%
\providecommand \natexlab [1]{#1}%
\providecommand \enquote  [1]{``#1''}%
\providecommand \bibnamefont  [1]{#1}%
\providecommand \bibfnamefont [1]{#1}%
\providecommand \citenamefont [1]{#1}%
\providecommand \href@noop [0]{\@secondoftwo}%
\providecommand \href [0]{\begingroup \@sanitize@url \@href}%
\providecommand \@href[1]{\@@startlink{#1}\@@href}%
\providecommand \@@href[1]{\endgroup#1\@@endlink}%
\providecommand \@sanitize@url [0]{\catcode `\\12\catcode `\$12\catcode
  `\&12\catcode `\#12\catcode `\^12\catcode `\_12\catcode `\%12\relax}%
\providecommand \@@startlink[1]{}%
\providecommand \@@endlink[0]{}%
\providecommand \url  [0]{\begingroup\@sanitize@url \@url }%
\providecommand \@url [1]{\endgroup\@href {#1}{\urlprefix }}%
\providecommand \urlprefix  [0]{URL }%
\providecommand \Eprint [0]{\href }%
\providecommand \doibase [0]{http://dx.doi.org/}%
\providecommand \selectlanguage [0]{\@gobble}%
\providecommand \bibinfo  [0]{\@secondoftwo}%
\providecommand \bibfield  [0]{\@secondoftwo}%
\providecommand \translation [1]{[#1]}%
\providecommand \BibitemOpen [0]{}%
\providecommand \bibitemStop [0]{}%
\providecommand \bibitemNoStop [0]{.\EOS\space}%
\providecommand \EOS [0]{\spacefactor3000\relax}%
\providecommand \BibitemShut  [1]{\csname bibitem#1\endcsname}%
\let\auto@bib@innerbib\@empty
\bibitem [{\citenamefont {Duong}\ \emph {et~al.}(2004)\citenamefont {Duong},
  \citenamefont {Satoh},\ and\ \citenamefont {Fiebig}}]{Duong2004}%
  \BibitemOpen
  \bibfield  {author} {\bibinfo {author} {\bibfnamefont {N.~P.}\ \bibnamefont
  {Duong}}, \bibinfo {author} {\bibfnamefont {T.}~\bibnamefont {Satoh}}, \ and\
  \bibinfo {author} {\bibfnamefont {M.}~\bibnamefont {Fiebig}},\ }\href
  {\doibase 10.1103/PhysRevLett.93.117402} {\bibfield  {journal} {\bibinfo
  {journal} {Phys. Rev. Lett.}\ }\textbf {\bibinfo {volume} {93}},\ \bibinfo
  {pages} {117402} (\bibinfo {year} {2004})}\BibitemShut {NoStop}%
\bibitem [{\citenamefont {Meier}\ \emph {et~al.}(2003)\citenamefont {Meier},
  \citenamefont {Levy},\ and\ \citenamefont {Loss}}]{Meier2003}%
  \BibitemOpen
  \bibfield  {author} {\bibinfo {author} {\bibfnamefont {F.}~\bibnamefont
  {Meier}}, \bibinfo {author} {\bibfnamefont {J.}~\bibnamefont {Levy}}, \ and\
  \bibinfo {author} {\bibfnamefont {D.}~\bibnamefont {Loss}},\ }\href {\doibase
  10.1103/PhysRevB.68.134417} {\bibfield  {journal} {\bibinfo  {journal}
  {Physical Review B}\ }\textbf {\bibinfo {volume} {68}},\ \bibinfo {pages}
  {134417} (\bibinfo {year} {2003})}\BibitemShut {NoStop}%
\bibitem [{\citenamefont {Jungwirth}\ \emph {et~al.}(2016)\citenamefont
  {Jungwirth}, \citenamefont {Marti}, \citenamefont {Wadley},\ and\
  \citenamefont {Wunderlich}}]{Jungwirth2016}%
  \BibitemOpen
  \bibfield  {author} {\bibinfo {author} {\bibfnamefont {T.}~\bibnamefont
  {Jungwirth}}, \bibinfo {author} {\bibfnamefont {X.}~\bibnamefont {Marti}},
  \bibinfo {author} {\bibfnamefont {P.}~\bibnamefont {Wadley}}, \ and\ \bibinfo
  {author} {\bibfnamefont {J.}~\bibnamefont {Wunderlich}},\ }\href {\doibase
  10.1038/nnano.2016.18} {\bibfield  {journal} {\bibinfo  {journal} {Nature
  Nanotechnology}\ }\textbf {\bibinfo {volume} {11}},\ \bibinfo {pages} {231}
  (\bibinfo {year} {2016})}\BibitemShut {NoStop}%
\bibitem [{\citenamefont {MacDonald}\ and\ \citenamefont
  {Tsoi}(2011)}]{MacDonald2011}%
  \BibitemOpen
  \bibfield  {author} {\bibinfo {author} {\bibfnamefont {A.~H.}\ \bibnamefont
  {MacDonald}}\ and\ \bibinfo {author} {\bibfnamefont {M.}~\bibnamefont
  {Tsoi}},\ }\href {\doibase 10.1098/rsta.2011.0014} {\bibfield  {journal}
  {\bibinfo  {journal} {Philosophical Transactions of the Royal Society A:
  Mathematical, Physical and Engineering Sciences}\ }\textbf {\bibinfo {volume}
  {369}},\ \bibinfo {pages} {3098} (\bibinfo {year} {2011})}\BibitemShut
  {NoStop}%
\bibitem [{\citenamefont {Gomonay}\ and\ \citenamefont
  {Loktev}(2014)}]{Gomonay2014}%
  \BibitemOpen
  \bibfield  {author} {\bibinfo {author} {\bibfnamefont {E.~V.}\ \bibnamefont
  {Gomonay}}\ and\ \bibinfo {author} {\bibfnamefont {V.~M.}\ \bibnamefont
  {Loktev}},\ }\href {\doibase 10.1063/1.4862467} {\bibfield  {journal}
  {\bibinfo  {journal} {Low Temperature Physics}\ }\textbf {\bibinfo {volume}
  {40}},\ \bibinfo {pages} {17} (\bibinfo {year} {2014})}\BibitemShut {NoStop}%
\bibitem [{\citenamefont {Gomonay}\ \emph {et~al.}(2016)\citenamefont
  {Gomonay}, \citenamefont {Jungwirth},\ and\ \citenamefont
  {Sinova}}]{Gomonay2016-06}%
  \BibitemOpen
  \bibfield  {author} {\bibinfo {author} {\bibfnamefont {O.}~\bibnamefont
  {Gomonay}}, \bibinfo {author} {\bibfnamefont {T.}~\bibnamefont {Jungwirth}},
  \ and\ \bibinfo {author} {\bibfnamefont {J.}~\bibnamefont {Sinova}},\ }\href
  {\doibase 10.1103/PhysRevLett.117.017202} {\bibfield  {journal} {\bibinfo
  {journal} {Phys. Rev. Lett.}\ }\textbf {\bibinfo {volume} {117}},\ \bibinfo
  {pages} {017202} (\bibinfo {year} {2016})}\BibitemShut {NoStop}%
\bibitem [{\citenamefont {Gomonay}\ and\ \citenamefont
  {Loktev}(2015)}]{Gomonay2015}%
  \BibitemOpen
  \bibfield  {author} {\bibinfo {author} {\bibfnamefont {O.~V.}\ \bibnamefont
  {Gomonay}}\ and\ \bibinfo {author} {\bibfnamefont {V.~M.}\ \bibnamefont
  {Loktev}},\ }\href {\doibase 10.1063/1.4931648} {\bibfield  {journal}
  {\bibinfo  {journal} {Low Temperature Physics}\ }\textbf {\bibinfo {volume}
  {41}},\ \bibinfo {pages} {698} (\bibinfo {year} {2015})}\BibitemShut
  {NoStop}%
\bibitem [{\citenamefont {Gomonay}\ \emph {et~al.}(2018)\citenamefont
  {Gomonay}, \citenamefont {Jungwirth},\ and\ \citenamefont
  {Sinova}}]{Gomonay2018sep}%
  \BibitemOpen
  \bibfield  {author} {\bibinfo {author} {\bibfnamefont {O.}~\bibnamefont
  {Gomonay}}, \bibinfo {author} {\bibfnamefont {T.}~\bibnamefont {Jungwirth}},
  \ and\ \bibinfo {author} {\bibfnamefont {J.}~\bibnamefont {Sinova}},\ }\href
  {\doibase 10.1103/PhysRevB.98.104430} {\bibfield  {journal} {\bibinfo
  {journal} {Phys. Rev. B}\ }\textbf {\bibinfo {volume} {98}},\ \bibinfo
  {pages} {104430} (\bibinfo {year} {2018})}\BibitemShut {NoStop}%
\bibitem [{\citenamefont {Keffer}\ and\ \citenamefont
  {Kittel}(1952)}]{Keffer1952}%
  \BibitemOpen
  \bibfield  {author} {\bibinfo {author} {\bibfnamefont {F.}~\bibnamefont
  {Keffer}}\ and\ \bibinfo {author} {\bibfnamefont {C.}~\bibnamefont
  {Kittel}},\ }\href {\doibase 10.1103/PhysRev.85.329} {\bibfield  {journal}
  {\bibinfo  {journal} {Phys. Rev.}\ }\textbf {\bibinfo {volume} {85}},\
  \bibinfo {pages} {329} (\bibinfo {year} {1952})}\BibitemShut {NoStop}%
\bibitem [{\citenamefont {Sievers}\ and\ \citenamefont
  {Tinkham}(1963)}]{Sievers1963}%
  \BibitemOpen
  \bibfield  {author} {\bibinfo {author} {\bibfnamefont {A.~J.}\ \bibnamefont
  {Sievers}}\ and\ \bibinfo {author} {\bibfnamefont {M.}~\bibnamefont
  {Tinkham}},\ }\href {\doibase 10.1103/PhysRev.129.1566} {\bibfield  {journal}
  {\bibinfo  {journal} {Phys. Rev.}\ }\textbf {\bibinfo {volume} {129}},\
  \bibinfo {pages} {1566} (\bibinfo {year} {1963})}\BibitemShut {NoStop}%
\bibitem [{\citenamefont {Cheng}\ \emph {et~al.}(2015)\citenamefont {Cheng},
  \citenamefont {Daniels}, \citenamefont {Zhu},\ and\ \citenamefont
  {Xiao}}]{Cheng2015}%
  \BibitemOpen
  \bibfield  {author} {\bibinfo {author} {\bibfnamefont {R.}~\bibnamefont
  {Cheng}}, \bibinfo {author} {\bibfnamefont {M.~W.}\ \bibnamefont {Daniels}},
  \bibinfo {author} {\bibfnamefont {J.-G.}\ \bibnamefont {Zhu}}, \ and\
  \bibinfo {author} {\bibfnamefont {D.}~\bibnamefont {Xiao}},\ }\href {\doibase
  10.1103/PhysRevB.91.064423} {\bibfield  {journal} {\bibinfo  {journal} {Phys.
  Rev. B}\ }\textbf {\bibinfo {volume} {91}},\ \bibinfo {pages} {064423}
  (\bibinfo {year} {2015})}\BibitemShut {NoStop}%
\bibitem [{\citenamefont {Khymyn}\ \emph {et~al.}(2017)\citenamefont {Khymyn},
  \citenamefont {Lisenkov}, \citenamefont {Tiberkevich}, \citenamefont
  {Ivanov},\ and\ \citenamefont {Slavin}}]{Khymyn2017}%
  \BibitemOpen
  \bibfield  {author} {\bibinfo {author} {\bibfnamefont {R.}~\bibnamefont
  {Khymyn}}, \bibinfo {author} {\bibfnamefont {I.}~\bibnamefont {Lisenkov}},
  \bibinfo {author} {\bibfnamefont {V.}~\bibnamefont {Tiberkevich}}, \bibinfo
  {author} {\bibfnamefont {B.~A.}\ \bibnamefont {Ivanov}}, \ and\ \bibinfo
  {author} {\bibfnamefont {A.}~\bibnamefont {Slavin}},\ }\href
  {https://doi.org/10.1038/srep43705} {\bibfield  {journal} {\bibinfo
  {journal} {Scientific Reports}\ }\textbf {\bibinfo {volume} {7}},\ \bibinfo
  {pages} {43705} (\bibinfo {year} {2017})}\BibitemShut {NoStop}%
\bibitem [{\citenamefont {N\'u\~nez}\ \emph {et~al.}(2006)\citenamefont
  {N\'u\~nez}, \citenamefont {Duine}, \citenamefont {Haney},\ and\
  \citenamefont {MacDonald}}]{Nunez2006}%
  \BibitemOpen
  \bibfield  {author} {\bibinfo {author} {\bibfnamefont {A.~S.}\ \bibnamefont
  {N\'u\~nez}}, \bibinfo {author} {\bibfnamefont {R.~A.}\ \bibnamefont
  {Duine}}, \bibinfo {author} {\bibfnamefont {P.}~\bibnamefont {Haney}}, \ and\
  \bibinfo {author} {\bibfnamefont {A.~H.}\ \bibnamefont {MacDonald}},\ }\href
  {\doibase 10.1103/PhysRevB.73.214426} {\bibfield  {journal} {\bibinfo
  {journal} {Phys. Rev. B}\ }\textbf {\bibinfo {volume} {73}},\ \bibinfo
  {pages} {214426} (\bibinfo {year} {2006})}\BibitemShut {NoStop}%
\bibitem [{\citenamefont {Haney}\ \emph {et~al.}(2007)\citenamefont {Haney},
  \citenamefont {Waldron}, \citenamefont {Duine}, \citenamefont {N\'u\~nez},
  \citenamefont {Guo},\ and\ \citenamefont {MacDonald}}]{Haney2007}%
  \BibitemOpen
  \bibfield  {author} {\bibinfo {author} {\bibfnamefont {P.~M.}\ \bibnamefont
  {Haney}}, \bibinfo {author} {\bibfnamefont {D.}~\bibnamefont {Waldron}},
  \bibinfo {author} {\bibfnamefont {R.~A.}\ \bibnamefont {Duine}}, \bibinfo
  {author} {\bibfnamefont {A.~S.}\ \bibnamefont {N\'u\~nez}}, \bibinfo {author}
  {\bibfnamefont {H.}~\bibnamefont {Guo}}, \ and\ \bibinfo {author}
  {\bibfnamefont {A.~H.}\ \bibnamefont {MacDonald}},\ }\href {\doibase
  10.1103/PhysRevB.75.174428} {\bibfield  {journal} {\bibinfo  {journal} {Phys.
  Rev. B}\ }\textbf {\bibinfo {volume} {75}},\ \bibinfo {pages} {174428}
  (\bibinfo {year} {2007})}\BibitemShut {NoStop}%
\bibitem [{\citenamefont {Duine}\ \emph {et~al.}(2007)\citenamefont {Duine},
  \citenamefont {Haney}, \citenamefont {N\'u\~nez},\ and\ \citenamefont
  {MacDonald}}]{Duine2007}%
  \BibitemOpen
  \bibfield  {author} {\bibinfo {author} {\bibfnamefont {R.~A.}\ \bibnamefont
  {Duine}}, \bibinfo {author} {\bibfnamefont {P.~M.}\ \bibnamefont {Haney}},
  \bibinfo {author} {\bibfnamefont {A.~S.}\ \bibnamefont {N\'u\~nez}}, \ and\
  \bibinfo {author} {\bibfnamefont {A.~H.}\ \bibnamefont {MacDonald}},\ }\href
  {\doibase 10.1103/PhysRevB.75.014433} {\bibfield  {journal} {\bibinfo
  {journal} {Phys. Rev. B}\ }\textbf {\bibinfo {volume} {75}},\ \bibinfo
  {pages} {014433} (\bibinfo {year} {2007})}\BibitemShut {NoStop}%
\bibitem [{\citenamefont {Marrows}(2016)}]{Marrows2016}%
  \BibitemOpen
  \bibfield  {author} {\bibinfo {author} {\bibfnamefont {C.}~\bibnamefont
  {Marrows}},\ }\href {\doibase 10.1126/science.aad8211} {\bibfield  {journal}
  {\bibinfo  {journal} {Science}\ }\textbf {\bibinfo {volume} {351}},\ \bibinfo
  {pages} {558} (\bibinfo {year} {2016})}\BibitemShut {NoStop}%
\bibitem [{\citenamefont {Wadley}\ \emph {et~al.}(2016)\citenamefont {Wadley},
  \citenamefont {Howells}, \citenamefont {{\v Z}elezn{\'y}}, \citenamefont
  {Andrews}, \citenamefont {Hills}, \citenamefont {Campion}, \citenamefont
  {Nov{\'a}k}, \citenamefont {Olejn{\'\i}k}, \citenamefont {Maccherozzi},
  \citenamefont {Dhesi}, \citenamefont {Martin}, \citenamefont {Wagner},
  \citenamefont {Wunderlich}, \citenamefont {Freimuth}, \citenamefont
  {Mokrousov}, \citenamefont {Kune{\v s}}, \citenamefont {Chauhan},
  \citenamefont {Grzybowski}, \citenamefont {Rushforth}, \citenamefont
  {Edmonds}, \citenamefont {Gallagher},\ and\ \citenamefont
  {Jungwirth}}]{Wadley2016}%
  \BibitemOpen
  \bibfield  {author} {\bibinfo {author} {\bibfnamefont {P.}~\bibnamefont
  {Wadley}}, \bibinfo {author} {\bibfnamefont {B.}~\bibnamefont {Howells}},
  \bibinfo {author} {\bibfnamefont {J.}~\bibnamefont {{\v Z}elezn{\'y}}},
  \bibinfo {author} {\bibfnamefont {C.}~\bibnamefont {Andrews}}, \bibinfo
  {author} {\bibfnamefont {V.}~\bibnamefont {Hills}}, \bibinfo {author}
  {\bibfnamefont {R.~P.}\ \bibnamefont {Campion}}, \bibinfo {author}
  {\bibfnamefont {V.}~\bibnamefont {Nov{\'a}k}}, \bibinfo {author}
  {\bibfnamefont {K.}~\bibnamefont {Olejn{\'\i}k}}, \bibinfo {author}
  {\bibfnamefont {F.}~\bibnamefont {Maccherozzi}}, \bibinfo {author}
  {\bibfnamefont {S.~S.}\ \bibnamefont {Dhesi}}, \bibinfo {author}
  {\bibfnamefont {S.~Y.}\ \bibnamefont {Martin}}, \bibinfo {author}
  {\bibfnamefont {T.}~\bibnamefont {Wagner}}, \bibinfo {author} {\bibfnamefont
  {J.}~\bibnamefont {Wunderlich}}, \bibinfo {author} {\bibfnamefont
  {F.}~\bibnamefont {Freimuth}}, \bibinfo {author} {\bibfnamefont
  {Y.}~\bibnamefont {Mokrousov}}, \bibinfo {author} {\bibfnamefont
  {J.}~\bibnamefont {Kune{\v s}}}, \bibinfo {author} {\bibfnamefont {J.~S.}\
  \bibnamefont {Chauhan}}, \bibinfo {author} {\bibfnamefont {M.~J.}\
  \bibnamefont {Grzybowski}}, \bibinfo {author} {\bibfnamefont {A.~W.}\
  \bibnamefont {Rushforth}}, \bibinfo {author} {\bibfnamefont {K.~W.}\
  \bibnamefont {Edmonds}}, \bibinfo {author} {\bibfnamefont {B.~L.}\
  \bibnamefont {Gallagher}}, \ and\ \bibinfo {author} {\bibfnamefont
  {T.}~\bibnamefont {Jungwirth}},\ }\href {\doibase 10.1126/science.aab1031}
  {\bibfield  {journal} {\bibinfo  {journal} {Science}\ }\textbf {\bibinfo
  {volume} {351}},\ \bibinfo {pages} {587} (\bibinfo {year}
  {2016})}\BibitemShut {NoStop}%
\bibitem [{\citenamefont {Hahn}\ \emph {et~al.}(2014)\citenamefont {Hahn},
  \citenamefont {de~Loubens}, \citenamefont {Naletov}, \citenamefont {Youssef},
  \citenamefont {Klein},\ and\ \citenamefont {Viret}}]{Hahn2014}%
  \BibitemOpen
  \bibfield  {author} {\bibinfo {author} {\bibfnamefont {C.}~\bibnamefont
  {Hahn}}, \bibinfo {author} {\bibfnamefont {G.}~\bibnamefont {de~Loubens}},
  \bibinfo {author} {\bibfnamefont {V.~V.}\ \bibnamefont {Naletov}}, \bibinfo
  {author} {\bibfnamefont {J.~B.}\ \bibnamefont {Youssef}}, \bibinfo {author}
  {\bibfnamefont {O.}~\bibnamefont {Klein}}, \ and\ \bibinfo {author}
  {\bibfnamefont {M.}~\bibnamefont {Viret}},\ }\href {\doibase
  10.1209/0295-5075/108/57005} {\bibfield  {journal} {\bibinfo  {journal}
  {{EPL} (Europhysics Letters)}\ }\textbf {\bibinfo {volume} {108}},\ \bibinfo
  {pages} {57005} (\bibinfo {year} {2014})}\BibitemShut {NoStop}%
\bibitem [{\citenamefont {Wang}\ \emph {et~al.}(2015)\citenamefont {Wang},
  \citenamefont {Du}, \citenamefont {Hammel},\ and\ \citenamefont
  {Yang}}]{Wang2015}%
  \BibitemOpen
  \bibfield  {author} {\bibinfo {author} {\bibfnamefont {H.}~\bibnamefont
  {Wang}}, \bibinfo {author} {\bibfnamefont {C.}~\bibnamefont {Du}}, \bibinfo
  {author} {\bibfnamefont {P.~C.}\ \bibnamefont {Hammel}}, \ and\ \bibinfo
  {author} {\bibfnamefont {F.}~\bibnamefont {Yang}},\ }\href {\doibase
  10.1103/PhysRevB.91.220410} {\bibfield  {journal} {\bibinfo  {journal} {Phys.
  Rev. B}\ }\textbf {\bibinfo {volume} {91}},\ \bibinfo {pages} {220410(R)}
  (\bibinfo {year} {2015})}\BibitemShut {NoStop}%
\bibitem [{\citenamefont {Lebrun}\ \emph {et~al.}(2018)\citenamefont {Lebrun},
  \citenamefont {Ross}, \citenamefont {Bender}, \citenamefont {Qaiumzadeh},
  \citenamefont {Baldrati}, \citenamefont {Cramer}, \citenamefont {Brataas},
  \citenamefont {Duine},\ and\ \citenamefont {Kl{\"a}ui}}]{Lebrun2018}%
  \BibitemOpen
  \bibfield  {author} {\bibinfo {author} {\bibfnamefont {R.}~\bibnamefont
  {Lebrun}}, \bibinfo {author} {\bibfnamefont {A.}~\bibnamefont {Ross}},
  \bibinfo {author} {\bibfnamefont {S.~A.}\ \bibnamefont {Bender}}, \bibinfo
  {author} {\bibfnamefont {A.}~\bibnamefont {Qaiumzadeh}}, \bibinfo {author}
  {\bibfnamefont {L.}~\bibnamefont {Baldrati}}, \bibinfo {author}
  {\bibfnamefont {J.}~\bibnamefont {Cramer}}, \bibinfo {author} {\bibfnamefont
  {A.}~\bibnamefont {Brataas}}, \bibinfo {author} {\bibfnamefont {R.~A.}\
  \bibnamefont {Duine}}, \ and\ \bibinfo {author} {\bibfnamefont
  {M.}~\bibnamefont {Kl{\"a}ui}},\ }\href {\doibase 10.1038/s41586-018-0490-7}
  {\bibfield  {journal} {\bibinfo  {journal} {Nature}\ }\textbf {\bibinfo
  {volume} {561}},\ \bibinfo {pages} {222} (\bibinfo {year}
  {2018})}\BibitemShut {NoStop}%
\bibitem [{\citenamefont {Baldrati}\ \emph {et~al.}(2019)\citenamefont
  {Baldrati}, \citenamefont {Gomonay}, \citenamefont {Ross}, \citenamefont
  {Filianina}, \citenamefont {Lebrun}, \citenamefont {Ramos}, \citenamefont
  {Leveille}, \citenamefont {Fuhrmann}, \citenamefont {Forrest}, \citenamefont
  {Maccherozzi}, \citenamefont {Valencia}, \citenamefont {Kronast},
  \citenamefont {Saitoh}, \citenamefont {Sinova},\ and\ \citenamefont
  {Kl\"aui}}]{Baldrati2019}%
  \BibitemOpen
  \bibfield  {author} {\bibinfo {author} {\bibfnamefont {L.}~\bibnamefont
  {Baldrati}}, \bibinfo {author} {\bibfnamefont {O.}~\bibnamefont {Gomonay}},
  \bibinfo {author} {\bibfnamefont {A.}~\bibnamefont {Ross}}, \bibinfo {author}
  {\bibfnamefont {M.}~\bibnamefont {Filianina}}, \bibinfo {author}
  {\bibfnamefont {R.}~\bibnamefont {Lebrun}}, \bibinfo {author} {\bibfnamefont
  {R.}~\bibnamefont {Ramos}}, \bibinfo {author} {\bibfnamefont
  {C.}~\bibnamefont {Leveille}}, \bibinfo {author} {\bibfnamefont
  {F.}~\bibnamefont {Fuhrmann}}, \bibinfo {author} {\bibfnamefont {T.~R.}\
  \bibnamefont {Forrest}}, \bibinfo {author} {\bibfnamefont {F.}~\bibnamefont
  {Maccherozzi}}, \bibinfo {author} {\bibfnamefont {S.}~\bibnamefont
  {Valencia}}, \bibinfo {author} {\bibfnamefont {F.}~\bibnamefont {Kronast}},
  \bibinfo {author} {\bibfnamefont {E.}~\bibnamefont {Saitoh}}, \bibinfo
  {author} {\bibfnamefont {J.}~\bibnamefont {Sinova}}, \ and\ \bibinfo {author}
  {\bibfnamefont {M.}~\bibnamefont {Kl\"aui}},\ }\href {\doibase
  10.1103/PhysRevLett.123.177201} {\bibfield  {journal} {\bibinfo  {journal}
  {Phys. Rev. Lett.}\ }\textbf {\bibinfo {volume} {123}},\ \bibinfo {pages}
  {177201} (\bibinfo {year} {2019})}\BibitemShut {NoStop}%
\bibitem [{\citenamefont {Hutchings}\ and\ \citenamefont
  {Samuelsen}(1972)}]{Hutchings1972}%
  \BibitemOpen
  \bibfield  {author} {\bibinfo {author} {\bibfnamefont {M.~T.}\ \bibnamefont
  {Hutchings}}\ and\ \bibinfo {author} {\bibfnamefont {E.~J.}\ \bibnamefont
  {Samuelsen}},\ }\href {\doibase 10.1103/PhysRevB.6.3447} {\bibfield
  {journal} {\bibinfo  {journal} {Phys. Rev. B}\ }\textbf {\bibinfo {volume}
  {6}},\ \bibinfo {pages} {3447} (\bibinfo {year} {1972})}\BibitemShut
  {NoStop}%
\bibitem [{\citenamefont {Mondal}\ \emph {et~al.}(2019)\citenamefont {Mondal},
  \citenamefont {Donges}, \citenamefont {Ritzmann}, \citenamefont {Oppeneer},\
  and\ \citenamefont {Nowak}}]{Mondal2019}%
  \BibitemOpen
  \bibfield  {author} {\bibinfo {author} {\bibfnamefont {R.}~\bibnamefont
  {Mondal}}, \bibinfo {author} {\bibfnamefont {A.}~\bibnamefont {Donges}},
  \bibinfo {author} {\bibfnamefont {U.}~\bibnamefont {Ritzmann}}, \bibinfo
  {author} {\bibfnamefont {P.~M.}\ \bibnamefont {Oppeneer}}, \ and\ \bibinfo
  {author} {\bibfnamefont {U.}~\bibnamefont {Nowak}},\ }\href {\doibase
  10.1103/PhysRevB.100.060409} {\bibfield  {journal} {\bibinfo  {journal}
  {Phys. Rev. B}\ }\textbf {\bibinfo {volume} {100}},\ \bibinfo {pages}
  {060409(R)} (\bibinfo {year} {2019})}\BibitemShut {NoStop}%
\bibitem [{\citenamefont {Uchida}\ \emph {et~al.}(1967)\citenamefont {Uchida},
  \citenamefont {Fukuoka}, \citenamefont {Kondoh}, \citenamefont {Takeda},
  \citenamefont {Nakazumi},\ and\ \citenamefont {Nagamiya}}]{Uchida1967}%
  \BibitemOpen
  \bibfield  {author} {\bibinfo {author} {\bibfnamefont {E.}~\bibnamefont
  {Uchida}}, \bibinfo {author} {\bibfnamefont {N.}~\bibnamefont {Fukuoka}},
  \bibinfo {author} {\bibfnamefont {H.}~\bibnamefont {Kondoh}}, \bibinfo
  {author} {\bibfnamefont {T.}~\bibnamefont {Takeda}}, \bibinfo {author}
  {\bibfnamefont {Y.}~\bibnamefont {Nakazumi}}, \ and\ \bibinfo {author}
  {\bibfnamefont {T.}~\bibnamefont {Nagamiya}},\ }\href {\doibase
  10.1143/JPSJ.23.1197} {\bibfield  {journal} {\bibinfo  {journal} {Journal of
  the Physical Society of Japan}\ }\textbf {\bibinfo {volume} {23}},\ \bibinfo
  {pages} {1197} (\bibinfo {year} {1967})}\BibitemShut {NoStop}%
\bibitem [{\citenamefont {Baltz}\ \emph {et~al.}(2018)\citenamefont {Baltz},
  \citenamefont {Manchon}, \citenamefont {Tsoi}, \citenamefont {Moriyama},
  \citenamefont {Ono},\ and\ \citenamefont {Tserkovnyak}}]{Baltz2018}%
  \BibitemOpen
  \bibfield  {author} {\bibinfo {author} {\bibfnamefont {V.}~\bibnamefont
  {Baltz}}, \bibinfo {author} {\bibfnamefont {A.}~\bibnamefont {Manchon}},
  \bibinfo {author} {\bibfnamefont {M.}~\bibnamefont {Tsoi}}, \bibinfo {author}
  {\bibfnamefont {T.}~\bibnamefont {Moriyama}}, \bibinfo {author}
  {\bibfnamefont {T.}~\bibnamefont {Ono}}, \ and\ \bibinfo {author}
  {\bibfnamefont {Y.}~\bibnamefont {Tserkovnyak}},\ }\href {\doibase
  10.1103/RevModPhys.90.015005} {\bibfield  {journal} {\bibinfo  {journal}
  {Rev. Mod. Phys.}\ }\textbf {\bibinfo {volume} {90}},\ \bibinfo {pages}
  {015005} (\bibinfo {year} {2018})}\BibitemShut {NoStop}%
\bibitem [{\citenamefont {Kampfrath}\ \emph {et~al.}(2010)\citenamefont
  {Kampfrath}, \citenamefont {Sell}, \citenamefont {Klatt}, \citenamefont
  {Pashkin}, \citenamefont {M{\"a}hrlein}, \citenamefont {Dekorsy},
  \citenamefont {Wolf}, \citenamefont {Fiebig}, \citenamefont {Leitenstorfer},\
  and\ \citenamefont {Huber}}]{Kampfrath2010}%
  \BibitemOpen
  \bibfield  {author} {\bibinfo {author} {\bibfnamefont {T.}~\bibnamefont
  {Kampfrath}}, \bibinfo {author} {\bibfnamefont {A.}~\bibnamefont {Sell}},
  \bibinfo {author} {\bibfnamefont {G.}~\bibnamefont {Klatt}}, \bibinfo
  {author} {\bibfnamefont {A.}~\bibnamefont {Pashkin}}, \bibinfo {author}
  {\bibfnamefont {S.}~\bibnamefont {M{\"a}hrlein}}, \bibinfo {author}
  {\bibfnamefont {T.}~\bibnamefont {Dekorsy}}, \bibinfo {author} {\bibfnamefont
  {M.}~\bibnamefont {Wolf}}, \bibinfo {author} {\bibfnamefont {M.}~\bibnamefont
  {Fiebig}}, \bibinfo {author} {\bibfnamefont {A.}~\bibnamefont
  {Leitenstorfer}}, \ and\ \bibinfo {author} {\bibfnamefont {R.}~\bibnamefont
  {Huber}},\ }\href {https://doi.org/10.1038/nphoton.2010.259} {\bibfield
  {journal} {\bibinfo  {journal} {Nature Photonics}\ }\textbf {\bibinfo
  {volume} {5}},\ \bibinfo {pages} {31} (\bibinfo {year} {2010})}\BibitemShut
  {NoStop}%
\bibitem [{\citenamefont {Kampfrath}\ \emph {et~al.}(2013)\citenamefont
  {Kampfrath}, \citenamefont {Battiato}, \citenamefont {Maldonado},
  \citenamefont {Eilers}, \citenamefont {N{\"o}tzold}, \citenamefont
  {M{\"a}hrlein}, \citenamefont {Zbarsky}, \citenamefont {Freimuth},
  \citenamefont {Mokrousov}, \citenamefont {Bl{\"u}gel}, \citenamefont {Wolf},
  \citenamefont {Radu}, \citenamefont {Oppeneer},\ and\ \citenamefont
  {M{\"u}nzenberg}}]{Kampfrath2013}%
  \BibitemOpen
  \bibfield  {author} {\bibinfo {author} {\bibfnamefont {T.}~\bibnamefont
  {Kampfrath}}, \bibinfo {author} {\bibfnamefont {M.}~\bibnamefont {Battiato}},
  \bibinfo {author} {\bibfnamefont {P.}~\bibnamefont {Maldonado}}, \bibinfo
  {author} {\bibfnamefont {G.}~\bibnamefont {Eilers}}, \bibinfo {author}
  {\bibfnamefont {J.}~\bibnamefont {N{\"o}tzold}}, \bibinfo {author}
  {\bibfnamefont {S.}~\bibnamefont {M{\"a}hrlein}}, \bibinfo {author}
  {\bibfnamefont {V.}~\bibnamefont {Zbarsky}}, \bibinfo {author} {\bibfnamefont
  {F.}~\bibnamefont {Freimuth}}, \bibinfo {author} {\bibfnamefont
  {Y.}~\bibnamefont {Mokrousov}}, \bibinfo {author} {\bibfnamefont
  {S.}~\bibnamefont {Bl{\"u}gel}}, \bibinfo {author} {\bibfnamefont
  {M.}~\bibnamefont {Wolf}}, \bibinfo {author} {\bibfnamefont {I.}~\bibnamefont
  {Radu}}, \bibinfo {author} {\bibfnamefont {P.~M.}\ \bibnamefont {Oppeneer}},
  \ and\ \bibinfo {author} {\bibfnamefont {M.}~\bibnamefont {M{\"u}nzenberg}},\
  }\href {https://doi.org/10.1038/nnano.2013.43} {\bibfield  {journal}
  {\bibinfo  {journal} {Nature Nanotechnology}\ }\textbf {\bibinfo {volume}
  {8}},\ \bibinfo {pages} {256} (\bibinfo {year} {2013})}\BibitemShut {NoStop}%
\bibitem [{\citenamefont {Bogdanov}\ and\ \citenamefont
  {Dragunov}(1998)}]{Bogdanov1998}%
  \BibitemOpen
  \bibfield  {author} {\bibinfo {author} {\bibfnamefont {A.~N.}\ \bibnamefont
  {Bogdanov}}\ and\ \bibinfo {author} {\bibfnamefont {I.~E.}\ \bibnamefont
  {Dragunov}},\ }\href {\doibase 10.1063/1.593515} {\bibfield  {journal}
  {\bibinfo  {journal} {Low Temperature Physics}\ }\textbf {\bibinfo {volume}
  {24}},\ \bibinfo {pages} {852} (\bibinfo {year} {1998})}\BibitemShut
  {NoStop}%
\bibitem [{\citenamefont {Skomski}\ \emph {et~al.}(2008)\citenamefont {Skomski}
  \emph {et~al.}}]{Skomski2008}%
  \BibitemOpen
  \bibfield  {author} {\bibinfo {author} {\bibfnamefont {R.}~\bibnamefont
  {Skomski}} \emph {et~al.},\ }\href@noop {} {\emph {\bibinfo {title} {Simple
  models of magnetism}}}\ (\bibinfo  {publisher} {Oxford University Press on
  Demand},\ \bibinfo {year} {2008})\BibitemShut {NoStop}%
\bibitem [{\citenamefont {Satoh}\ \emph {et~al.}(2010)\citenamefont {Satoh},
  \citenamefont {Cho}, \citenamefont {Iida}, \citenamefont {Shimura},
  \citenamefont {Kuroda}, \citenamefont {Ueda}, \citenamefont {Ueda},
  \citenamefont {Ivanov}, \citenamefont {Nori},\ and\ \citenamefont
  {Fiebig}}]{Satoh2010}%
  \BibitemOpen
  \bibfield  {author} {\bibinfo {author} {\bibfnamefont {T.}~\bibnamefont
  {Satoh}}, \bibinfo {author} {\bibfnamefont {S.-J.}\ \bibnamefont {Cho}},
  \bibinfo {author} {\bibfnamefont {R.}~\bibnamefont {Iida}}, \bibinfo {author}
  {\bibfnamefont {T.}~\bibnamefont {Shimura}}, \bibinfo {author} {\bibfnamefont
  {K.}~\bibnamefont {Kuroda}}, \bibinfo {author} {\bibfnamefont
  {H.}~\bibnamefont {Ueda}}, \bibinfo {author} {\bibfnamefont {Y.}~\bibnamefont
  {Ueda}}, \bibinfo {author} {\bibfnamefont {B.~A.}\ \bibnamefont {Ivanov}},
  \bibinfo {author} {\bibfnamefont {F.}~\bibnamefont {Nori}}, \ and\ \bibinfo
  {author} {\bibfnamefont {M.}~\bibnamefont {Fiebig}},\ }\href {\doibase
  10.1103/PhysRevLett.105.077402} {\bibfield  {journal} {\bibinfo  {journal}
  {Phys. Rev. Lett.}\ }\textbf {\bibinfo {volume} {105}},\ \bibinfo {pages}
  {077402} (\bibinfo {year} {2010})}\BibitemShut {NoStop}%
\bibitem [{\citenamefont {Baierl}\ \emph {et~al.}(2016)\citenamefont {Baierl},
  \citenamefont {Mentink}, \citenamefont {Hohenleutner}, \citenamefont {Braun},
  \citenamefont {Do}, \citenamefont {Lange}, \citenamefont {Sell},
  \citenamefont {Fiebig}, \citenamefont {Woltersdorf}, \citenamefont
  {Kampfrath},\ and\ \citenamefont {Huber}}]{Baierl2016}%
  \BibitemOpen
  \bibfield  {author} {\bibinfo {author} {\bibfnamefont {S.}~\bibnamefont
  {Baierl}}, \bibinfo {author} {\bibfnamefont {J.~H.}\ \bibnamefont {Mentink}},
  \bibinfo {author} {\bibfnamefont {M.}~\bibnamefont {Hohenleutner}}, \bibinfo
  {author} {\bibfnamefont {L.}~\bibnamefont {Braun}}, \bibinfo {author}
  {\bibfnamefont {T.-M.}\ \bibnamefont {Do}}, \bibinfo {author} {\bibfnamefont
  {C.}~\bibnamefont {Lange}}, \bibinfo {author} {\bibfnamefont
  {A.}~\bibnamefont {Sell}}, \bibinfo {author} {\bibfnamefont {M.}~\bibnamefont
  {Fiebig}}, \bibinfo {author} {\bibfnamefont {G.}~\bibnamefont {Woltersdorf}},
  \bibinfo {author} {\bibfnamefont {T.}~\bibnamefont {Kampfrath}}, \ and\
  \bibinfo {author} {\bibfnamefont {R.}~\bibnamefont {Huber}},\ }\href
  {\doibase 10.1103/PhysRevLett.117.197201} {\bibfield  {journal} {\bibinfo
  {journal} {Phys. Rev. Lett.}\ }\textbf {\bibinfo {volume} {117}},\ \bibinfo
  {pages} {197201} (\bibinfo {year} {2016})}\BibitemShut {NoStop}%
\bibitem [{\citenamefont {{Kohmoto}}\ and\ \citenamefont
  {{Moriyasu}}(2018)}]{Kohmoto2018}%
  \BibitemOpen
  \bibfield  {author} {\bibinfo {author} {\bibfnamefont {T.}~\bibnamefont
  {{Kohmoto}}}\ and\ \bibinfo {author} {\bibfnamefont {T.}~\bibnamefont
  {{Moriyasu}}},\ }in\ \href {\doibase 10.1109/IRMMW-THz.2018.8510270} {\emph
  {\bibinfo {booktitle} {2018 43rd International Conference on Infrared,
  Millimeter, and Terahertz Waves (IRMMW-THz)}}}\ (\bibinfo {year} {2018})\
  pp.\ \bibinfo {pages} {1--2}\BibitemShut {NoStop}%
\bibitem [{\citenamefont {Milano}\ \emph {et~al.}(2004)\citenamefont {Milano},
  \citenamefont {Steren},\ and\ \citenamefont {Grimsditch}}]{Milano2004}%
  \BibitemOpen
  \bibfield  {author} {\bibinfo {author} {\bibfnamefont {J.}~\bibnamefont
  {Milano}}, \bibinfo {author} {\bibfnamefont {L.~B.}\ \bibnamefont {Steren}},
  \ and\ \bibinfo {author} {\bibfnamefont {M.}~\bibnamefont {Grimsditch}},\
  }\href {\doibase 10.1103/PhysRevLett.93.077601} {\bibfield  {journal}
  {\bibinfo  {journal} {Phys. Rev. Lett.}\ }\textbf {\bibinfo {volume} {93}},\
  \bibinfo {pages} {077601} (\bibinfo {year} {2004})}\BibitemShut {NoStop}%
\bibitem [{\citenamefont {Kurosawa}\ \emph {et~al.}(1980)\citenamefont
  {Kurosawa}, \citenamefont {Miura},\ and\ \citenamefont
  {Saito}}]{Kurosawa1980}%
  \BibitemOpen
  \bibfield  {author} {\bibinfo {author} {\bibfnamefont {K.}~\bibnamefont
  {Kurosawa}}, \bibinfo {author} {\bibfnamefont {M.}~\bibnamefont {Miura}}, \
  and\ \bibinfo {author} {\bibfnamefont {S.}~\bibnamefont {Saito}},\ }\href
  {\doibase 10.1088/0022-3719/13/8/021} {\bibfield  {journal} {\bibinfo
  {journal} {Journal of Physics C: Solid State Physics}\ }\textbf {\bibinfo
  {volume} {13}},\ \bibinfo {pages} {1521} (\bibinfo {year}
  {1980})}\BibitemShut {NoStop}%
\bibitem [{\citenamefont {Chikazumi}(1997)}]{Chikazumi1997}%
  \BibitemOpen
  \bibfield  {author} {\bibinfo {author} {\bibfnamefont {S.}~\bibnamefont
  {Chikazumi}},\ }\href@noop {} {{\selectlanguage {English}\emph {\bibinfo
  {title} {Physics of {{Ferromagnetism}}}}}},\ International {{Series}} of
  {{Monographs}} on {{Physics}}\ (\bibinfo  {publisher} {{Oxford Science
  Publications}},\ \bibinfo {year} {1997})\BibitemShut {NoStop}%
\bibitem [{\citenamefont {Gomonay}\ and\ \citenamefont
  {Loktev}(2010)}]{Gomonay2010}%
  \BibitemOpen
  \bibfield  {author} {\bibinfo {author} {\bibfnamefont {H.~V.}\ \bibnamefont
  {Gomonay}}\ and\ \bibinfo {author} {\bibfnamefont {V.~M.}\ \bibnamefont
  {Loktev}},\ }\href {\doibase 10.1103/PhysRevB.81.144427} {\bibfield
  {journal} {\bibinfo  {journal} {Phys. Rev. B}\ }\textbf {\bibinfo {volume}
  {81}},\ \bibinfo {pages} {144427} (\bibinfo {year} {2010})}\BibitemShut
  {NoStop}%
\bibitem [{\citenamefont {Cheng}\ \emph {et~al.}(2014)\citenamefont {Cheng},
  \citenamefont {Xiao}, \citenamefont {Niu},\ and\ \citenamefont
  {Brataas}}]{Cheng2014}%
  \BibitemOpen
  \bibfield  {author} {\bibinfo {author} {\bibfnamefont {R.}~\bibnamefont
  {Cheng}}, \bibinfo {author} {\bibfnamefont {J.}~\bibnamefont {Xiao}},
  \bibinfo {author} {\bibfnamefont {Q.}~\bibnamefont {Niu}}, \ and\ \bibinfo
  {author} {\bibfnamefont {A.}~\bibnamefont {Brataas}},\ }\href {\doibase
  10.1103/PhysRevLett.113.057601} {\bibfield  {journal} {\bibinfo  {journal}
  {Phys. Rev. Lett.}\ }\textbf {\bibinfo {volume} {113}},\ \bibinfo {pages}
  {057601} (\bibinfo {year} {2014})}\BibitemShut {NoStop}%
\bibitem [{\citenamefont {Tranchida}\ \emph {et~al.}(2018)\citenamefont
  {Tranchida}, \citenamefont {Plimpton}, \citenamefont {Thibaudeau},\ and\
  \citenamefont {Thompson}}]{Tranchida2018}%
  \BibitemOpen
  \bibfield  {author} {\bibinfo {author} {\bibfnamefont {J.}~\bibnamefont
  {Tranchida}}, \bibinfo {author} {\bibfnamefont {S.}~\bibnamefont {Plimpton}},
  \bibinfo {author} {\bibfnamefont {P.}~\bibnamefont {Thibaudeau}}, \ and\
  \bibinfo {author} {\bibfnamefont {A.}~\bibnamefont {Thompson}},\ }\href
  {\doibase https://doi.org/10.1016/j.jcp.2018.06.042} {\bibfield  {journal}
  {\bibinfo  {journal} {Journal of Computational Physics}\ }\textbf {\bibinfo
  {volume} {372}},\ \bibinfo {pages} {406 } (\bibinfo {year}
  {2018})}\BibitemShut {NoStop}%
\bibitem [{\citenamefont {Vansteenkiste}\ \emph {et~al.}(2014)\citenamefont
  {Vansteenkiste}, \citenamefont {Leliaert}, \citenamefont {Dvornik},
  \citenamefont {Helsen}, \citenamefont {Garcia-Sanchez},\ and\ \citenamefont
  {Van~Waeyenberge}}]{Vansteenkiste2014}%
  \BibitemOpen
  \bibfield  {author} {\bibinfo {author} {\bibfnamefont {A.}~\bibnamefont
  {Vansteenkiste}}, \bibinfo {author} {\bibfnamefont {J.}~\bibnamefont
  {Leliaert}}, \bibinfo {author} {\bibfnamefont {M.}~\bibnamefont {Dvornik}},
  \bibinfo {author} {\bibfnamefont {M.}~\bibnamefont {Helsen}}, \bibinfo
  {author} {\bibfnamefont {F.}~\bibnamefont {Garcia-Sanchez}}, \ and\ \bibinfo
  {author} {\bibfnamefont {B.}~\bibnamefont {Van~Waeyenberge}},\ }\href
  {\doibase 10.1063/1.4899186} {\bibfield  {journal} {\bibinfo  {journal} {AIP
  Advances}\ }\textbf {\bibinfo {volume} {4}},\ \bibinfo {pages} {107133}
  (\bibinfo {year} {2014})}\BibitemShut {NoStop}%
\bibitem [{\citenamefont {Slonczewski}(1996)}]{Slonczewski1996}%
  \BibitemOpen
  \bibfield  {author} {\bibinfo {author} {\bibfnamefont {J.}~\bibnamefont
  {Slonczewski}},\ }\href {\doibase
  https://doi.org/10.1016/0304-8853(96)00062-5} {\bibfield  {journal} {\bibinfo
   {journal} {Journal of Magnetism and Magnetic Materials}\ }\textbf {\bibinfo
  {volume} {159}},\ \bibinfo {pages} {L1 } (\bibinfo {year}
  {1996})}\BibitemShut {NoStop}%
\bibitem [{\citenamefont {Kamra}\ \emph {et~al.}(2018)\citenamefont {Kamra},
  \citenamefont {Troncoso}, \citenamefont {Belzig},\ and\ \citenamefont
  {Brataas}}]{Kamra2018}%
  \BibitemOpen
  \bibfield  {author} {\bibinfo {author} {\bibfnamefont {A.}~\bibnamefont
  {Kamra}}, \bibinfo {author} {\bibfnamefont {R.~E.}\ \bibnamefont {Troncoso}},
  \bibinfo {author} {\bibfnamefont {W.}~\bibnamefont {Belzig}}, \ and\ \bibinfo
  {author} {\bibfnamefont {A.}~\bibnamefont {Brataas}},\ }\href {\doibase
  10.1103/PhysRevB.98.184402} {\bibfield  {journal} {\bibinfo  {journal} {Phys.
  Rev. B}\ }\textbf {\bibinfo {volume} {98}},\ \bibinfo {pages} {184402}
  (\bibinfo {year} {2018})}\BibitemShut {NoStop}%
\bibitem [{\citenamefont {Kimel}\ \emph {et~al.}(2009)\citenamefont {Kimel},
  \citenamefont {Ivanov}, \citenamefont {Pisarev}, \citenamefont {Usachev},
  \citenamefont {Kirilyuk},\ and\ \citenamefont {Rasing}}]{Kimel2009}%
  \BibitemOpen
  \bibfield  {author} {\bibinfo {author} {\bibfnamefont {A.~V.}\ \bibnamefont
  {Kimel}}, \bibinfo {author} {\bibfnamefont {B.~A.}\ \bibnamefont {Ivanov}},
  \bibinfo {author} {\bibfnamefont {R.~V.}\ \bibnamefont {Pisarev}}, \bibinfo
  {author} {\bibfnamefont {P.~A.}\ \bibnamefont {Usachev}}, \bibinfo {author}
  {\bibfnamefont {A.}~\bibnamefont {Kirilyuk}}, \ and\ \bibinfo {author}
  {\bibfnamefont {T.}~\bibnamefont {Rasing}},\ }\href {\doibase
  10.1038/nphys1369} {\bibfield  {journal} {\bibinfo  {journal} {Nature
  Physics}\ }\textbf {\bibinfo {volume} {5}},\ \bibinfo {pages} {727} (\bibinfo
  {year} {2009})}\BibitemShut {NoStop}%
\bibitem [{\citenamefont {Chen}\ \emph {et~al.}(2018)\citenamefont {Chen},
  \citenamefont {Zarzuela}, \citenamefont {Zhang}, \citenamefont {Song},
  \citenamefont {Zhou}, \citenamefont {Shi}, \citenamefont {Li}, \citenamefont
  {Zhou}, \citenamefont {Jiang}, \citenamefont {Pan},\ and\ \citenamefont
  {Tserkovnyak}}]{Chen2018}%
  \BibitemOpen
  \bibfield  {author} {\bibinfo {author} {\bibfnamefont {X.~Z.}\ \bibnamefont
  {Chen}}, \bibinfo {author} {\bibfnamefont {R.}~\bibnamefont {Zarzuela}},
  \bibinfo {author} {\bibfnamefont {J.}~\bibnamefont {Zhang}}, \bibinfo
  {author} {\bibfnamefont {C.}~\bibnamefont {Song}}, \bibinfo {author}
  {\bibfnamefont {X.~F.}\ \bibnamefont {Zhou}}, \bibinfo {author}
  {\bibfnamefont {G.~Y.}\ \bibnamefont {Shi}}, \bibinfo {author} {\bibfnamefont
  {F.}~\bibnamefont {Li}}, \bibinfo {author} {\bibfnamefont {H.~A.}\
  \bibnamefont {Zhou}}, \bibinfo {author} {\bibfnamefont {W.~J.}\ \bibnamefont
  {Jiang}}, \bibinfo {author} {\bibfnamefont {F.}~\bibnamefont {Pan}}, \ and\
  \bibinfo {author} {\bibfnamefont {Y.}~\bibnamefont {Tserkovnyak}},\ }\href
  {\doibase 10.1103/PhysRevLett.120.207204} {\bibfield  {journal} {\bibinfo
  {journal} {Phys. Rev. Lett.}\ }\textbf {\bibinfo {volume} {120}},\ \bibinfo
  {pages} {207204} (\bibinfo {year} {2018})}\BibitemShut {NoStop}%
\bibitem [{\citenamefont {Nussle}\ \emph {et~al.}(2019)\citenamefont {Nussle},
  \citenamefont {Thibaudeau},\ and\ \citenamefont {Nicolis}}]{Nussle2019}%
  \BibitemOpen
  \bibfield  {author} {\bibinfo {author} {\bibfnamefont {T.}~\bibnamefont
  {Nussle}}, \bibinfo {author} {\bibfnamefont {P.}~\bibnamefont {Thibaudeau}},
  \ and\ \bibinfo {author} {\bibfnamefont {S.}~\bibnamefont {Nicolis}},\ }\href
  {\doibase 10.1103/PhysRevB.100.214428} {\bibfield  {journal} {\bibinfo
  {journal} {Phys. Rev. B}\ }\textbf {\bibinfo {volume} {100}},\ \bibinfo
  {pages} {214428} (\bibinfo {year} {2019})},\ \Eprint
  {http://arxiv.org/abs/1907.01857} {arXiv:1907.01857} \BibitemShut {NoStop}%
\bibitem [{\citenamefont {Yamane}\ \emph {et~al.}(2019)\citenamefont {Yamane},
  \citenamefont {Gomonay},\ and\ \citenamefont {Sinova}}]{Yamane2019}%
  \BibitemOpen
  \bibfield  {author} {\bibinfo {author} {\bibfnamefont {Y.}~\bibnamefont
  {Yamane}}, \bibinfo {author} {\bibfnamefont {O.}~\bibnamefont {Gomonay}}, \
  and\ \bibinfo {author} {\bibfnamefont {J.}~\bibnamefont {Sinova}},\ }\href
  {\doibase 10.1103/PhysRevB.100.054415} {\bibfield  {journal} {\bibinfo
  {journal} {Phys. Rev. B}\ }\textbf {\bibinfo {volume} {100}},\ \bibinfo
  {pages} {054415} (\bibinfo {year} {2019})}\BibitemShut {NoStop}%
\end{thebibliography}%

\end{document}